\documentclass{jaa}

\def\k{{\bf k}}

\def\HI{{\rm HI}}

\def\physrep{PHYS REP}
 \def\prd{PhReD}

\def\mnras{MNRAS}
\def\aaps{AAPS}
\def\aap{AAP}
\def\apj{Ap.J}
\def\aj{A.J}

\def\apjl{Ap.JL}
\def\u{{\bf U}} 
\def\d{\vec{d}}
\def\HI{{\rm HI \,}}
\def\th{\vec{\theta}}
\def\V{{\mathcal V}}
\def\S{\mathcal{S}}
\def\N{{\mathcal N}}
\def\F{\mathcal{F}}

\def\del{\partial}

\def\rn{r_{\nu}}
\def\rnp{r_{\nu}^{'}}

\def\k{{\bf k}}

\def\kpr{{\bf k}_\perp}

\def\lsim{~\rlap{$<$}{\lower 1.0ex\hbox{$\sim$}}}

\def\gsim{~\rlap{$>$}{\lower 1.0ex\hbox{$\sim$}}}

\usepackage{epsfig}
\begin{document}
\title[HI signal from Post-reionization era with the ORT]
{Prospects for detecting  the    $326.5 \,{\rm MHz}$
redshifted $21 \,{\rm cm}$ HI signal  with the Ooty Radio Telescope (ORT)} 
\author[ S. S. Ali and S. Bharadwaj ] { Sk. Saiyad
Ali$^{1}$\thanks{Email:saiyad@phys.jdvu.ac.in}  and
Somnath Bharadwaj$^{2}$\thanks{Email:somnath@cts.iitkgp.ernet.in} \\ 
$^{1}$ Department of Physics,Jadavpur University, Kolkata 700032, India\\
 $^{2}$ Department of Physics and Meteorology \& Centre for
  Theoretical Studies , IIT Kharagpur, 721 302 , India }
\date {}
\maketitle
\begin{abstract} 
Observations of the redshifted $21\, \rm cm$ HI fluctuations promise
to be an important probe of the post-reionization era ($z \le \,
6$). In this paper we  calculate the 
expected  signal and foregrounds for the  upgraded Ooty Radio Telescope
(ORT) which operates   at frequency $\nu_{o}=326.5 \, {\rm 
  MHz}$ which corresponds to  redshift $z=3.35 $. Assuming that the 
visibilities  contain only the HI signal and system 
noise, we show that a 3$\sigma$ detection of  the HI signal ($\sim
1 \,{\rm mK}$) is possible at angular scales $11^{'}$ to
$3^{\circ}$ with $\approx \,1000$ hours  of observation.   
 Foreground removal is one of the major challenges for a statistical detection of the
redshifted $21\,\rm cm$ HI signal.  We 
assess the contribution of different foregrounds and find that  the $326.5
\,\rm{ MHz}$ sky is dominated by  the extragalactic point sources at
the angular scales of our interest.  The expected total
foregrounds are $ 10^4 - 10^5$ times higher than the HI signal.

\end{abstract}

\section{Introduction}

The study of the evolution of  cosmic structure has been an important
subject in cosmology. In the post reionization era $(z<6)$ the
    21-cm emission originates 
from    dense pockets of self-shielded hydrogen.  These systems which are
seen as   Damped Lyman-$\alpha$   absorption lines (DLAs) in quasar
spectra   are known to contain the bulk of the HI (Zafar et
al. 2013). Different from traditional  galaxy redshift surveys, 21cm
surveys do not need to resolve individual HI sources. The collective
emission from  the individual clouds appears as a very faint,  diffuse
background radiation in all low frequency radio observations below $1420 \,{\rm
  MHz}$, and the source  clustering  is imprinted on the fluctuations
of 
this background radiation (Bharadwaj, Nath \& Sethi 2001). Observations of
the redshifted $21\,\rm cm$ radiation from neutral hydrogen (HI) 
 can in principle be carried out over a large redshift range
starting from the cosmological Dark Ages through the Epoch of
Reionization (EoR) to the present epoch. This allows us to study both the
evolution history of neutral hydrogen as well as the growth of large
scale structures in the Universe (Kumar, Padmanabhan \& Subramanian 1995;
Bagla, Nath \& Padmanabhan 1997; Madau, Meiksin \& Rees 1997;
Bharadwaj, Nath \& Sethi 2001;  Bharadwaj \& Pandey
2003; Bharadwaj \& Ali 2005; Furlanetto et
al. 2006; Wyithe \& Loeb  2008; Bagla, Khandai \& Datta
2010). Redshifted $21\,\rm cm$ observations also hold the
potential of  probing the expansion history of the
Universe (Visbal, Loeb \& Wyithe 2009;
Bharadwaj, Sethi \& Saini 2009). It has been
proposed that the Baryon Acoustic Oscillation (BAO) in the redshifted
$21\,\rm cm$ signal from the post-reionization era ($z \le 6)$ is a
very sensitive probe of dark energy (Wyithe,
Loeb \& Geil 2007; Chang et al. 2008; Seo et al. 2010; Masui et
al. 2010). A compact interferometer with a wide fields of view is 
needed to cover the BAO length-scale. By scanning across frequency, $21\,\rm
cm$ observations will probe the HI distribution at different times in
cosmic history. It will allow us to construct $21\,\rm cm$ tomography of
the IGM. This tomography  may carry more useful information than any other
survey in cosmology (Madau, Meiksin \& Rees 1997; Loeb \& Zaldarriaga
2004; Loeb \& Wyithe 2008).

Realizing this  great potential,  a  large number  of the recent or upcoming
radio-interferometric experiments are aimed at measuring the HI
$21\,\rm cm$ signal at different redshifts from z $\sim
1\, {\rm to} \,12$. The Giant Meterwave Radio Telescope 
(GMRT\footnote{http://www.gmrt.ncra.tifr.res.in}; Swarup et al. 1991)   
is functioning at several bands in the frequency range $ 150 - 1420\,{\rm
  MHz}$ and can potentially detect the $21\,{\rm cm}$ signal at high
 as well as low redshifts (Bharadwaj \& Ali 2005).Several
 low-frequency EoR experiments (LOFAR\footnote{http://www.lofar.org/}, 
  MWA\footnote{http://web.haystack.mit.edu/ast/arrays/MWA/}, 21CMA,
 formerly known as PAST\footnote{http://web.phys.cmu.edu/~past/},
 PAPER (Parsons et 
 al. 2010), LWA \footnote{http://lwa.nrl.navy.mil/}) are currently in
 progress or under construction. They have raised the possibility to
 detect and characterize the EoR signal. Several other upcoming radio
 telescopes like CHIME\footnote{http://chime.phas.ubc.ca/} and 
 BAOBAB\footnote{http://bao.berkeley.edu/} aim to  
probe the low redshift Universe ($ z \le 2.5$). It has been 
recently reported  that a cylindrical transit interferometer would be a novel
approach which would  avoid the curved sky complications of
conventional interferometry and be well suited  for 
wide-field observations (Shaw et al. 2013). They claim that the
data analysis techniques and two point statistics allow new ways of
tackling the important problems like map-making and foreground
removal. More ambitious designs are being planned for the future
 low frequency  telescope 
SKA\footnote{http://www.skatelescope.org/}. This would be well suited
for carrying out observations towards detecting the HI signal over
alarge  
redshift range $z \sim 0$ to $\sim 12 $.

 The removal of  continuum
foregrounds sources (such as extragalactic point sources, Galactic
synchrotron,  and Galactic  and extra-Galactic free free
emission) is a  major challenge for detecting the faint HI signal. The
foreground sources are expected to be roughly four to five orders of magnitude
stronger than the cosmological HI signal (Di Matteo et al. 2002; Ali,
Bharadwaj \& Chengalur 2008; Ghosh et al. 2011a). Various proposals
for tackling the foreground issue have been discussed in the
literature (Harkar et al. 2009; Bowman et al. 2009; Datta et al. 2010; 
Jelic et al. 2010; Bernardi et al. 2011; Ghosh et al. 2011b; Liu \&
Tegmark 2011; Mao 2012; Liu \& Tegmark 2012; Cho et al. 2012; Switzer
et al. 2013; Jacobs et al. 2013; Pober et al. 2013; Dillon et
al. 2013). The polarized Galactic Synchrotron emission  is expected to be 
Faraday-rotated along the path, and it may acquire additional spectral  structure 
through polarization leakage at the telescope.  This is a potential complication for 
detecting the HI signal.  The effect of polarized foregrounds on foreground 
removal has been studied by Moore et al. (2013).

 A statistical detection of the post-reionization HI signal has
already been made (Pen et al. 2009) through cross-correlation between
the HIPASS and the 6dfGRS. In a recent paper, Masui et al. (2013)
 have  measured the cross power spectrum at redshift $z \sim 0.8$ 
 using 21 cm intensity maps acquired at the Green Bank Telescope
 (GBT) and large-scale structure traced by optically selected
 galaxies in the WiggleZ Dark Energy Survey. This measurement puts
 a lower limit on the fluctuation power of 21 cm emission. For
the first time, Switzer et al. (2013) 
have measure the auto-power spectrum of redshifted $21 \,{\rm cm}$
radiation  from  the HI distribution at redshift $z \,\sim 0.8$   with
GBT. These detections represent important steps towards using
redshifted $21\,{\rm cm}$ surface brightness fluctuations to probe the
HI distribution at high $z$. 

 Efforts are currently underway (Prasad \& Subrahmanya 2011a,
  2011b)  
 to upgrade the Ooyt Radio Telescope  (hereafter ORT) so that it may
 be operated as a radio-interferometric array. 
The aim of this paper is to  present  the expected post-reionization  
 $21\,\rm cm$ signal at frequency $\nu_{o}=326.5 \, {\rm MHz}$
$(z=3.35)$, and discusses the possibility of its detection with the 
upgraded ORT.  For detecting this
faint cosmological signal, it is very crucial to understand all
foreground components in detailed. Here we use a  foreground model to 
predict the foreground contribution to the radio background at  $326.5
\, {\rm MHz}$.  The prospect for detecting the redshifted $21\,\rm
cm$ signal is considerably higher at at this frequency in comparison 
to the lower frequencies (e.g. $150 \, {\rm MHz}$, EoR)
where the foreground contribution and the system noise are both
larger.  

The  background UV radiation at redshift $(z=3.35)$
is expected to be nearly uniform, and we  expect the 
 redshifted $21\,\rm cm$ power spectrum to  trace the underlying
 matter power spectrum with a possible linear bias.   The ORT holds
 the potential of measuring the $z=3.35$ power spectrum, opening the
 possibility of probing large-scale structure formation at an hitherto
 unexplored redshift.  We note that it is extremely difficult to 
 accurately  measure  the redshift for a  large numbers of 
 galaxies at high redshifts (Eisenstein et al. 2005), and it will be
 difficult to probe $z>3$  using galaxy surveys. Further, 
the quasar distribution   is known to peak between $z = 2$ and 3
(Busca et al. 2013), and we do not expect   Lyman-$\alpha$  forest
surveys to be very effective at $z>3$. Observations of the redshifted
21-cm signal are possibly one of the few (if not only) techniques  by
which it will be possible to probe the matter power spectrum at $z>3$. 
This has the possibility of probing  cosmology and structure formation
through a variety of effects including the redshift space distortion 
(Bharadwaj etal. 2001; Bharadwaj \& Ali  2004; Barkana \& Loeb  2005;
Ali, Bharadwaj, \& Pandey 2005; Wang \& Hu 2006; Masui et al. 2010; Mao et al. 2012;
Majumdar et al. 2013 ) and the Alcock-Paczyński  test (Nusser 2005;
Barkana 2006).  Further, five successive  oscillations of the BAO are
well in the $k$  range that will be probed by ORT. The BAO is a
powerful probe of the expansion history, and a detection will
constrain   cosmological  parameters at $z=3.35$.    The present paper
is exploratory in nature, and it presents a preliminary estimate of
the expected signal and foregrounds. We plan to present more
quantitative estimates for parameter estimation in subsequent
publications.

A brief outline of the paper follows. Section $2$ introduces the
upgraded ORT as a radio interferometer and analyzes  the 
visibility signal that will be measured by this instrument. 
Section $3$  discusses how  the correlations between the measured
visibilities can be used  to quantify the angular and
frequency domain fluctuations of the background radiation. Section $4$ 
presents model prediction for the  HI signal, the signal to noise
ratio and the contributions from  different foregrounds components. 
This Section also discusses  the feasibility of detection of the
signal.  Section $5$ contains a 
summary and the conclusions.

In this work we have used the standard  LCDM cosmology with
parameters: $\Omega_{m0} = 0.30$, $\Omega_bh^2 = 0.024$,
$\Omega_{\Lambda 0} = 0.7$, $h = 0.7$, $n_s= 1.0$ and $\sigma_8 = 1.0$.

\section{ The ORT and the measured visibilities.}
The Ooty Radio Telescope (ORT) consists of a 530 m long and 30 m wide
parabolic cylindrical reflector. The telescope is placed in the
north-south direction on a hill with the same slope as the
latitude($11^{\circ}$) of the station (Swarup et al. 1971; Sarma et
al. 1975). It thus becomes possible to observe the same part of the
sky by rotating the parabolic cylinder along it’s long axis. The
telescope operates at a nominal frequency of $\nu_o=326.5 \, {\rm
  MHz}$ with $\lambda_o \, = \, 0.919 \, {\rm m}$.  The entire
telescope feed consists of 1056 half-wavelength ($0.5 \, \lambda_o
\approx 0.5 \, {\rm m}$) dipoles which are placed nearly end to end
along the focal line of the cylinder. The separation between the
centers of two successive dipoles is $ 0.515 \, \lambda_o$ which is
slightly larger than the length of each dipole. The entire feed is
placed off-axis to avoid maximally the obstruction of the incoming
radiation.

\begin{figure}
\begin{centering} 
\epsfig{scale=0.3,file=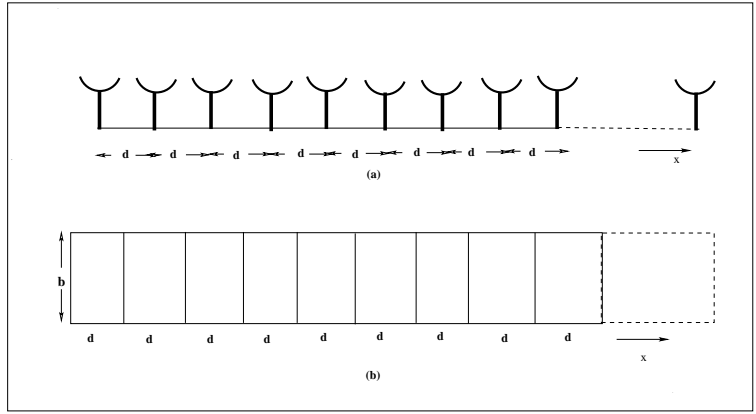}
\caption{ This  shows the antenna layout corresponding to the   ORT
  when it is used as a radio-interferometer.  We have a linear array 
  of antenna elements  with spacing $d$ arranged along the $x$ axis
  which is   aligned  to the axis of the cylindrical reflector. The
  figure also   shows the $b \times d$ rectangular aperture of the
  individual antenna   elements.  Here $b$ corresponds to the width of
  the parabolic cylindrical reflector. } 
\label{fig:antennad.eps} 
  \end{centering}
\end{figure}

 Work is currently underway to upgrade the ORT whereby the linear
 dipole array may be operated as a radio-interferometer. Here the
 signal from groups of dipoles is combined to form an antenna element.
 The RF signal from each antenna element is directly digitized and
 transported to a central location where the signals from different
 pairs of antenna elements are correlated to produce the visibilities
 $\V(\u,\nu)$ which are recorded.  Here $\u=\d/\lambda$ refers to a
 baseline which is the antenna separation (Figure
 \ref{fig:antennad.eps}) $\d$ in units of the observing wavelength
 $\lambda$.  The upgrade is being carried out in two different stages
 with two nearly independent systems, namely {\bf{Phase I}} and
 {\bf{Phase II}}, being expected at the end of the upgrade (Prasad \&
 Subrahmanya 2011a, 2011b).  We 
 briefly discuss these two phases below, and the relevant parameters
 are presented in Table~\ref{tab:array} (C. R. Subrahmanya, private
 communication). 

\noindent{\bf Phase I} Here $24$ successive dipoles are combined to
form a single antenna element.  This gives $40$ antennas each of which
is $11.5 \, {\rm m} $ along the length of the cylinder and $ 30 \,
{\rm m}$ wide. The smallest baseline corresponds to an antenna
separation of $11.5 \, {\rm m}$ and the longest baseline corresponds
to $448.5 \, {\rm m}$. The system has a frequency bandwidth of $18 \,
{\rm MHz}$.

\noindent{\bf Phase II} Here $4$ successive dipoles are combined to
form a single antenna element.  This gives $264$ antennas each of
which is $1.9 \, {\rm m} $ along the length of the cylinder and $ 30
\, {\rm m}$ wide. The smallest baseline also corresponds to an antenna
separation of $1.9 \, {\rm m}$ and the longest baseline corresponds to
$505.0 \, {\rm m}$. The system has a frequency bandwidth of $30 \,
{\rm MHz}$.

We note that CHIME, an upcoming new  telescope designed to detect
    the BAO,  is partly
    similar to the ORT in construction. 
 The      CHIME consists of five parabolic cylindrical
    reflectors,  ecah 
    of dimensions  100m $\times$ 20m and  each containing  
    $256$ antennas. The total telescope  is $100 \, {\rm m}
    \times 100\, {\rm m}$ in dimension. Unlike the ORT, this will be  a  drift scan
    telescope with no 
    moving     parts, and  it will cover the  frequency range $800 \,
    {\rm to}\,     400\, {\rm MHz}$ which corresponds to the redshift
    range $\sim 0.8$ to 
    $2.5$.

\begin{table}
\caption{System parameters for Phases I and II of the upgraded ORT.  }
\vspace{.2in}
\label{tab:array}
\begin{tabular}{|l|c|c|}
\hline \hline Parameter & Phase I &Phase II\\ \hline No. of antennas
($N_A$) & 40 & 264\\ \hline Aperture dimensions ($b \times d$) & $30
\,{\rm m} \times 11.5 \,{\rm m}$ & $ 30 \,{\rm m} \times 1.92 \,{\rm
  m} $ \\ \hline Field of View(FoV) & $ 1.75^{\circ} \times
4.6^{\circ}$ & $ 1.75^{\circ} \times 27.4^{\circ}$ \\ \hline Smallest
baseline ($d_{min}$) & $11.5 \,{\rm m} $ & $1.9 \,{\rm m} $\\ \hline
Largest baseline ($d_{max}$) & $ 448.5 \,{\rm m}$ & $505.0 \,{\rm
  m}$\\ \hline Angular resolution & $ 7^{'}$ & $6.3^{'}$\\ \hline
Total bandwidth (B) & $18 \,{\rm MHz}$ & $30 \,{\rm MHz}$ \\ \hline
Single Visibility rms. noise ($\sigma$) & & \\ assuming $T_{sys}=150
\, {\rm K}, \eta=0.6$, & $1.12$ Jy & $6.69 $ Jy \\ $\Delta \nu_c=0.1
\, {\rm MHz},\Delta t=16 \, {\rm s}$ & & \\ \hline
\end{tabular}
\end{table}

Figure \ref{fig:antennad.eps} provides a schematic representation of
the ORT  when it is used a radio-interferometer. The parabolic
cylinder may be thought of   as a linear array of $N_A$ radio
antennas,  each antenna located at a separation  $d$ along the length
of the cylinder.  Viewed from the direction in which the telescope
is pointing, each antenna has a rectangular aperture of 
dimensions $b \times d$ where $b=30 \, {\rm m}$ is the width of the
parabola, and $d= 11.5$ and $1.9$m for Phases I and II respectively. 
For convenience, we have assumed  the telescope aperture to lie in the
$x-y$ plane with the $x$ axis along the length of the cylinder. We
then have the baselines 
\begin{equation}
\u_1= \left( \frac{d}{\lambda}\right) \hat{i}; \hspace{0.5cm} \u_2= 2
\u_1;  \hspace{0.5cm} \u_3= 3 \u_1; ... \hspace{0.5cm}
 \u_{N_A-1}=(N_A-1) \u_1
\label{eq:a0}
\end{equation}
for which the   complex    visibilities  $\V(\u,\nu)$ are recorded. It
should be noted that  there  
is considerable redundancy in this radio-interferometric  array  {\it
  ie.} there are many different  antenna pairs  
which correspond to the same baseline. Any  baseline $\u_n$ 
occurs  $M_n=(N_A-n) $ times in the array. In reality $\u_1,\u_2,...$
change as $\nu$ varies across the observing bandwidth. This is an
extremely important factor that needs to be considered in the actual
data analysis. However this is not  very significant for the signal
and foreground estimates  presented here, and   we  ignore this 
 for the purpose of the present  analysis, and
 hold $\u$ fixed at  the value  corresponding to $\lambda_o$.

The visibility  $\V(\u,\nu)$ recorded  at any baseline $\u$ is the
Fourier transform of the product of  $I(\th,\,\nu)$  which is 
the specific intensity
distribution on the sky  and  $A(\th, \nu)$ which is the primary beam
pattern or the normalized power pattern of the individual antenna. We
have    
\begin{equation}
{\V}(\u,\nu)= \int d^2 \th \,A(\th, \, \nu) \,  I(\th, \,\nu) \, e^{- i
2 \pi \u \cdot \th} \, . 
\label{eq:a1}
\end{equation}
where $\th$ is a two dimensional vector in the plane of the sky with
origin at the center of the field of view,  and the  beam pattern
$A(\th,\,\nu)$  quantifies how the   
individual antenna  responds to signals from
different directions in the sky.   We have assumed that the field of
view of the individual antennas is 
sufficiently small so that we may ignore the curvature of the sky and
treat the region of sky under observation as being flat. While such an
assumption is quite justified for Phase I,  the field of
view is quite large for Phase II and the flat sky assumption is not
strictly valid in this situation. We, however, expect our predictions
based on the flat sky assumption to provide a reasonable preliminary
estimate of the  signal expected for both Phases I and II.

\begin{figure}[t]
\begin{centering} 
\includegraphics[width=120mm,height=8.cm,angle=0.]{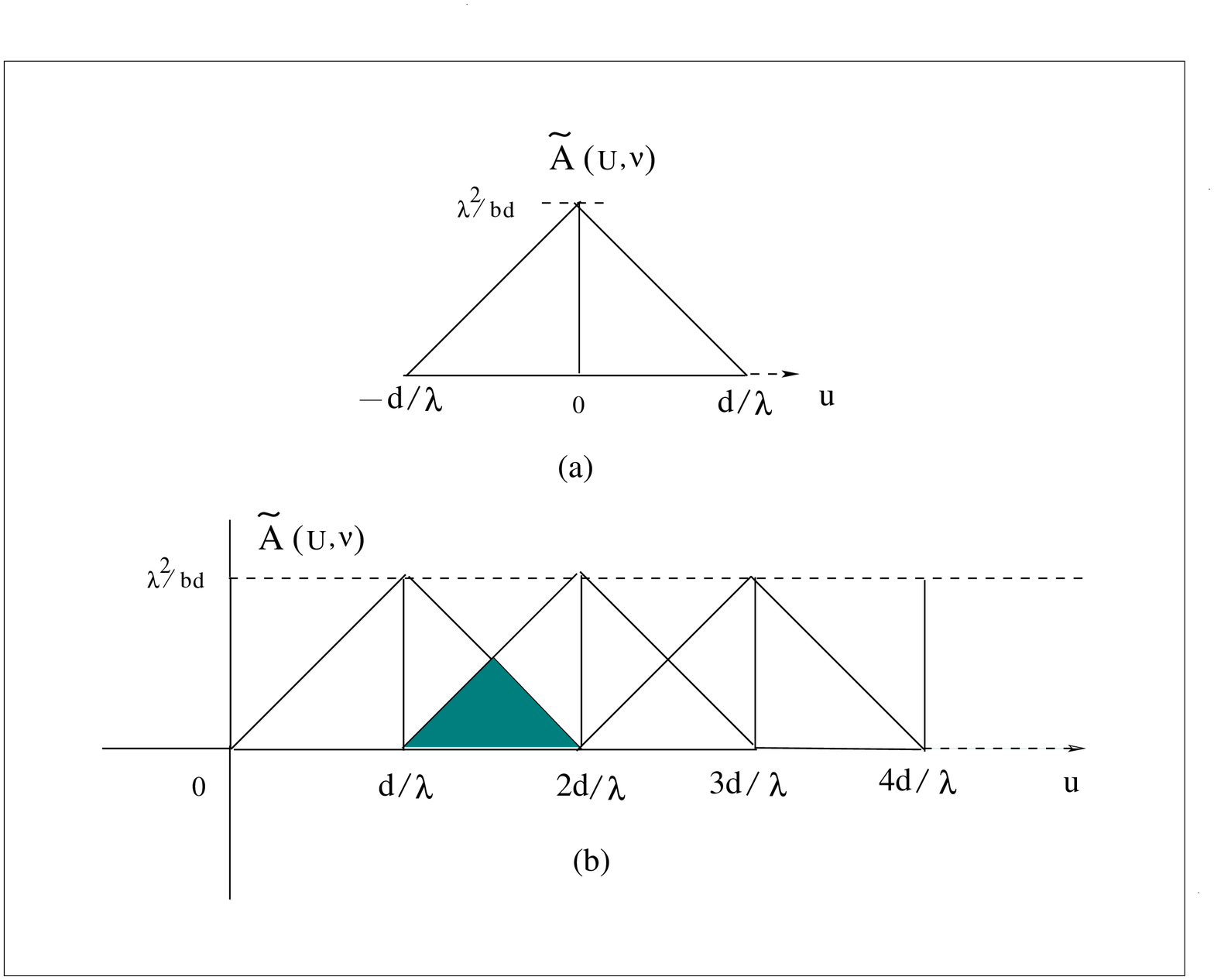}
\caption{The upper panel   shows a schematic view of the 
aperture power pattern $ \tilde{A}\,(\u, \nu)$  as a function of $u$
for $v=0$. The lower panel shows the $u$ range and the respective
weights corresponding to each Fourier mode that contributes to the
visibility at any baseline $\u_n$. The shaded region shows the overlap
between the Fourier modes that contribute to the visibilities at 
two adjacent  baselines.} 
\label{fig:af.eps} 
  \end{centering}
\end{figure}

We now briefly discuss   the normalized power pattern $A(\th,\nu)$ of
the individual  antennas in our radio-interferometer.  This can be
calculated by considering the
antenna as an emitter instead of receiver. We first calculate
$E(\th,\nu)$ the normalized far-field radiation electric  pattern
that will be  produced by the antenna, where $A(\th,\nu)=\mid
E(\th,\nu)\mid^2$. The function  $E(\th,\nu)$ is the
Fourier  transform of the  
electric field  pattern $\tilde{E}(\u,\nu)$ at the telescope's
aperture (Figure \ref{fig:antennad.eps}).
The ORT responds  only to a single polarization determined by
the dipole feeds which are aligned  parallel to the telescope
cylinder's axis. It is therefore  justified to  ignore the vector nature of the
electric field $\tilde{E}(\u,\nu)$ and focus on a single
polarization.  The exact form of $\tilde{E}(\u,\nu)$ depends  
on  how the dipole illuminates the  antenna aperture. 
Modeling this is quite complicated  and we do not attempt it here. For
the purpose of the present  analysis we  make the simplifying   
assumption  that the electric field $\tilde{E}(\u,\nu)$ is uniform
everywhere  on 
the  $b \times d$  rectangular aperture of the antenna element  (Figure  
\ref{fig:antennad.eps}) .   We then have   the primary beam pattern  
\begin{equation}
A(\th,\nu)=
{\rm sinc}^2\left(\frac{\pi\,d \,\theta_x}{\lambda}
\right)\,
{\rm sinc}^2\left(\frac{\pi\,b \,\theta_y}{\lambda}
\right)\,.
\label{eq:pbm}
\end{equation}
The actual beam pattern is expected to be somewhat broader than
predicted by eq. (\ref{eq:pbm})  if  the dipole's  illumination
pattern  is taken into account.   We use  the  aperture efficiency
$\eta$  that appears in the subsequent calculations to account for
this to some extent. 

The primary beam pattern decides the  field of view of the
radio-interferometer. We see that in this case (eq. \ref{eq:pbm})
we have an asymmetric field of view.  The primary beam has a
full width at half maximum (FWHM) of  $1.55^{\circ}$ corresponding to 
$b=30 \, {\rm m}$ in the east-west direction,  and  
$4.05^{\circ}$ and  $24.32^{\circ}$ in the north-south   direction for
Phases I and II respectively. The anisotropy in the field of view  is
particularly pronounced in Phase II where the N-S extent is more than
$10$ times the extent in the E-W direction.

It is useful to decompose the specific intensity as 
\begin{equation}
 I(\th,\nu)=\bar{I}(\nu)+\delta I(\th,\,\nu)
\end{equation}
where the first term is an  uniform  background brightness and the
second term is  the  angular fluctuation in the specific intensity. 
We use this and express eq. (\ref{eq:a1}) in terms of a convolution as  
\begin{equation}
{\V}(\u,\nu)=  \tilde{A}\,(\u, \nu)\,\bar{I}(\nu) + 
\tilde{A}\left(\u,\,\nu\right)\, \otimes \, \Delta
\tilde{I}(\u,\,\nu).   
\label{eq:a_1}
\end{equation}
where $\Delta \tilde{ I}(\u,\,\nu)$  and $\tilde{A}\,(\u, \nu)$
are the Fourier transforms of $\delta I(\th,\,\nu) $ and
$A(\th,\,\nu)$ respectively.  We refer to  $\tilde{A}\,(\u, \nu)$ as
the aperture power pattern. 

The aperture power pattern  $\tilde{A}\,(\u, \nu)$
 is  the auto-convolution of the electric field at the telescope
 aperture {\it ie.} $\tilde{A}(\u,\nu)=\tilde{E}(\u,\nu) \otimes
 \tilde{E}(\u,\nu)$. The 
telescope's aperture being finite, we have the interesting property
that $\tilde{A}(\u,\nu)$ has compact support in $\u$ irrespective of
the details of the shape of the telescope's aperture.  For the uniform
rectangular aperture assumed earlier we have 
\begin{equation}
\tilde{A}(\u, \nu)= \frac{\lambda^2}{bd} \,\Lambda \left(\frac{
  u\,\lambda}{ d}\, \right) \,\Lambda \, \left(\frac{v\,\lambda}{
  b}\,\right) \,.  
\label{eq:a2}
\end{equation}
where $\u=(u,v)$, and  $ \Lambda {(x)}$ is the triangular function
defined as 
\begin{equation}
\Lambda (x)=1-|x| \hspace{0.5cm}  {\rm  for}  \hspace{0.5cm}   |x| <
  1\,, \hspace{0.5cm}  {\rm and} \hspace{0.5cm} \Lambda
  (x)=0  \hspace{0.5cm}  {\rm 
    for}  \hspace{0.5cm}   |x| \ge 1 \,.
\label{eq:a3}
\end{equation}
We see that $\tilde{A}(\u,\nu)$  has non-zero values  only within   
$\mid u \mid < d/\lambda$,  $\mid v \mid < b/\lambda$ 
and vanished beyond.  Figure \ref{fig:af.eps}  shows the $u$
dependence of  $\tilde{A}(\u,\nu)$ for $v=0$. The actual behaviour of 
$\tilde{A}(\u,\nu)$ will be different, and we expect the $u,v$
dependence to 
fall faster than the the triangular function when the dipole's
illumination pattern is taken into account. However the fact that 
$\tilde{A}(\u,\nu)$ has compact support, and that 
$\tilde{A}(\u,\nu)=0$  for 
$\mid u \mid \ge d/\lambda$,  $\mid v \mid \ge b/\lambda$ 
will continue to hold.

We see that the first term $\tilde{A}\,(\u, \nu)\,\bar{I}(\nu)$ 
in eq. (\ref{eq:a_1}) dies away before the
smallest  baseline $\u_1$ (Figure  \ref{fig:af.eps}), and hence it is
not necessary to consider the contribution from $\bar{I}_{\nu}$.  We
then have  
\begin{equation}
{\V}(\u_n,\nu)=  \int \, d^2 U^{'}  \,
  \tilde{A}\left(\u_n-\u^{'},\,\nu\right)\, 
 \, \Delta \tilde{I}(\u{'},\,\nu).   
\label{eq:a4}
\end{equation}
where each visibility $\V(\u_{n},\nu)$ is a weighted  linear superposition of
of different Fourier modes $ \Delta\tilde{I}(\u,\,\nu)$. The
contribution peaks at $\u=\u_n$, and it is restricted to a rectangle of
size $(b/\lambda) \, \times   \, (d/\lambda)$ centered at $\u_n$. The 
modes outside this region do not contribute to $\V(\u_n,\nu)$.  This is
shown schematically in  Figure \ref{fig:af.eps}.  We also note that
there is overlap between the Fourier modes that contribute to two
neighbouring visibilities $\V(\u_{n},\nu)$ and $\V(\u_{n+1},\nu)$.
This implies that to some extent the same information is present in 
the visibilities measured at  two neighbouring baselines.  
This overlap, however, is restricted to the nearest neighbours, and
does not extend beyond.  

In addition to the sky signal discussed till now, each visibility also
has a noise contribution {\it ie.}
\begin{equation}
\V(\u_n,\nu)=\V^{\rm sky}(\u_n,\nu)+\N(\u_n,\nu)
\label{eq:a5}
\end{equation}
where the noise contribution $\N(\u_n,\nu)$  in each visibility is an
independent complex Gaussian random variable with zero mean.  The real 
part (or equivalent to  the imaginary part) of $\N(\u_n,\nu)$ has 
a rms. fluctuation (Thompson, Moran \& Swenson 1986) 
\begin{equation}
\sigma
=\frac{\sqrt{2}k_BT_{sys}}{\eta A\sqrt{\Delta \nu_c
      \Delta t}}
\label{eq:rms}
\end{equation}
where $T_{sys}$ is the total system temperature, $k_B$ is the Boltzmann
constant, $A=b \times d$ is the physical collecting area of each 
antenna, $\eta$ is the aperture efficiency, $\Delta \nu_c$ is the
channel width and $\Delta t$ is the correlator integration time. 
We expect $T_{sys}$ to have a value around $150 \, {\rm K}$, and we
use this value for the estimates presented here.  
Considering observations with $\Delta \nu_c=0.1 \, {\rm MHz}$ and $
\Delta t = 16 \, {\rm s}$  we have  $\sigma= 1.12 \, {\rm Jy}$ and
$6.69 \, {\rm   Jy}$ for   Phases I and II respectively. 

We now highlight two interesting features which are unique to the ORT
radio-interferometer. First, in a typical radio-interferometer the
baseline $\u$ corresponding to a pair of antennas changes with the
rotation of the Earth. As a consequence, the individual baselines sweep
out different tracks in the $u-v$ plane during the course of a long
observation. However, for the ORT the North-South axis of the
cylindrical reflector is parallel to the Earth's rotation axis.  The
baselines (eq. \ref{eq:a0}) too are all parallel to the Earth's
rotation  axis and they  do not change with the rotation of the
Earth. Second, the 
interferometer has a high degree of redundancy in that there are
$N_A-n$ distinct  antenna pairs which correspond to any particular
baseline  $\u_n$.  Considering a particular baseline $\u_m$, the
visibility $\V^{'}_{ab}$ measured by any    antenna pair   $a,b$ is
the actual  visibility $\V(\u_m)$ amplified by   the unknown
individual   antenna gains $g_a$ and $g_b$ {\it ie.} $\V^{'}_{ab}=g_a
\, g^*_b \V(\u)$.  The fact that there are many different antenna
pairs for which the measured visibility has the same signal $\V(\u_m)$
can be put to good use in determining the unknown antenna gains $g_a$
and $g_b$ (Ram Marthi \& Chengalur 2013). 

\section{Visibility correlations}
We assume that the observed sky signal $\delta I(\th,\,\nu) $  is a
particular realization of a statistically homogeneous and
isotropic random process.  In other words,  the process that generates 
$\delta I(\th,\,\nu) $ has no
preferred origin or direction on the sky. Further, 
it  also has   no preferred origin in frequency.  
We use the multi-frequency angular power spectrum $C_{\ell}(\Delta 
\nu)$ to quantify the  statistical properties of $\delta I(\th,\,\nu)
$ (Datta, Roy Choudhury \& Bharadwaj 2007).  The calculations are 
considerably simplified in the 
flat-sky approximation  where it is convenient to 
use Fourier modes instead of the spherical harmonics
$Y_{\ell}^m(\theta,\phi)$.  The two-dimensional power 
spectrum  $P(U,\Delta \nu)$ is defined  through  
\begin{equation}
  \langle \Delta {\tilde I}(\u,\,\nu) \,\Delta \tilde{
    I}^*(\u^{\prime},\,\nu +\Delta \nu)\rangle=  P(U,\Delta \nu)
\delta_D^2 \left(\u-\u^{\prime}\right) ,  
\label{eq:c1}
\end{equation}
where   $\delta_D^2 \left(\u-\u^{\prime}\right)$ is the two-dimensional
Dirac-delta function.
The angular brackets $\langle ... \rangle$ above denote the  ensemble 
average with respect to different realizations of $\delta I(\th,\,\nu)
$.  

The multi-frequency angular power spectrum $C_{\ell}(\Delta \nu)$
refers to $\delta T(\th,\,\nu)$ which is the brightness temperature 
corresponding to   $\delta I(\th,\,\nu)$.  We have 
\begin{equation}
C_{\ell}(\Delta\nu)=\left( \frac{\partial B}{\partial T}\right)^{-2}
P(\ell/2 \pi,\Delta \nu) \,.
\label{eq:c2}
\end{equation}  
where the angular multipole $\ell$ corresponds to the 2D dimensional
wave vector $2 \pi \u$ with $\ell=2 \pi \mid \u \mid$, 
$B$ is the Planck function and $({\partial B}/{\partial T})=2
k_B/\lambda^2$ in the Raleigh-Jeans limit which is valid at the
frequencies of our interest.  Strictly speaking,  we should evaluate
$({\partial B}/{\partial T})$ at two different frequencies   $\nu$ and
$\nu + \Delta \nu$.This introduces  a   slow variation of order
$\sim \Delta \nu/\nu$ which is  small in most  of our
analysis. For the  estimates of this paper, 
here and in the subsequent analysis, we ignore several such  terms
which introduce  slow variations of the order of 
$\sim \Delta \nu/\nu \ll 1$.

The multi-frequency angular power spectrum 
$C_{\ell}(\Delta \nu)$ defined above jointly characterizes 
the  angular ($\ell$) and frequency ($\Delta \nu$) dependence of the
statistical properties of the sky signal. 
The observed visibilities $\V(\u_n,\nu)$ are related to
$C_{\ell}(\Delta \nu)$ through the two visibility correlation 
which, using eqs. (\ref{eq:a4}), (\ref{eq:c1}) and (\ref{eq:c2}),
 can be written as
\begin{eqnarray}
\langle {\V}(\u_n, \nu)
\,{\V}^{*}(\u_m,\nu +\Delta \nu)\rangle&=& 
\left( \frac{\partial
  B}{\partial T}\right)^{2} 
 \int \, d^2 U^{'}  \, 
  \tilde{A}\left(\u_n-\u^{'},\nu\right)  \times \, \nonumber  \\
 &&  \tilde{A}^{*}\left(\u_m-\u^{'},\nu + \Delta \nu \right)\, 
C_{2 \pi U^{'}}(\Delta \nu) \,.
\label{eq:c3}
\end{eqnarray} 
The functions $  \tilde{A}\left(\u_n-\u^{'},\nu\right)$ and 
$\tilde{A}^{*}\left(\u_m-\u^{'},\nu + \Delta \nu \right)$  have an
overlap  only when $\mid n-m \mid \le 1$ (Figure
\ref{fig:antennad.eps}). This implies that two visibilities are
correlated only if they correspond to the same baseline or the nearest
neighbours. The correlation is strongest when $n=m$, and we restrict
our analysis here to this situation where the two baselines are the
same.  Further, in the subsequent discussion we also ignore the slow
$\Delta \nu$ dependence of  
$\tilde{A}^{*}\left(\u_m-\u^{'},\nu + \Delta \nu \right)$.  The two
visibility correlation can then be expressed as 
\begin{equation}
{V}_2(\u_n, \Delta \nu)  \equiv \langle {\V}(\u_n, \nu)
{\V}^{*}(\u_n,\nu +\Delta \nu)\rangle
\label{eq:c4}
\end{equation}
and we have 
\begin{eqnarray} 
 {V}_2(\u_n,\Delta \nu) =
\left( \frac{\partial  B}{\partial T}\right)^{2} 
\int d^2 {\u}^{\prime} \,|\tilde{A}\left(\u_n - 
 {\u}^{\prime} \right)|^2\, C_{2\,\pi\, U^{\prime}}(\Delta \nu). 
\label{eq:c5}    
\end{eqnarray}
where we do not explicitly show $\nu$ as an argument in any of the
terms,  and it is implicit that this has the value 
$\nu_o=326.5 \, {\rm MHz}$. 

At large baselines it is possible to approximate the convolution
in (eq. \ref{eq:c5}) as 
\begin{eqnarray} 
 {V}_2(\u_n,\Delta \nu) =
\left( \frac{\partial  B}{\partial T}\right)^{2} 
\left[ \int d^2 {\u}^{\prime} \,|\tilde{A}\left(\u_n -  
 {\u}^{\prime} \right)|^2\, \right] C_{\ell}(\Delta \nu). 
\label{eq:c6}    
\end{eqnarray}
with $\ell = 2 \pi U_n$. This gives a very simple relation 
\begin{eqnarray} 
 C_{\ell}(\Delta \nu) = 0.26 \, \left(\frac{{\rm mK}}{{\rm Jy}} 
 \right)^2 \, \left(\frac{b \  d}{{\rm m^2}} \right) \, 
 {V}_2(\u_n,\Delta \nu)
\label{eq:c7}    
\end{eqnarray}
between $ {V}_2(\u_n,\Delta \nu)$ which can be determined directly
from the measured visibilities and $C_{\ell}(\Delta \nu)$  which
quantifies the statistical properties of the brightness temperature
distribution on the sky. We expect (eq. \ref{eq:c7}) to be a good
approximation only at large baselines where the value of $C_{2 \pi \u}(\Delta \nu)$ does not  change much   within the width of the
function $|\tilde{A}\left(\u_n -  {\u} \right)|^2 $. 
However,  our
investigations later in this paper show that eq. \ref{eq:c7} provides
a reasonable good approximation to the full convolution
(eq. \ref{eq:c5}) for nearly the entire $U$ range covered by ORT. 

\section{Predictions}
The contribution to the measured visibilities  $\V(\u,\nu)$ 
from the sky signal (eq. \ref{eq:a5}) is 
  a combination of two  different components 
\begin{equation}
{\V}^{sky}(\u_n,\nu)={\S}(\u_n,\nu)+  {\F}(\u_n,\nu) \,.
\label{eq:v}
\end{equation}
where ${\S}(\u_n,\nu)$  is the HI signal which is the object of our
study here,  and 
${\F}(\u_n,\nu)$  is the contribution from other astrophysical sources
referred to as the foregrounds. We treat both of these, as well as the
system noise ${\N}(\u_n,\nu)$ as  uncorrelated random variables with
zero mean. We then have  
\begin{equation}
{V}_2(\u_n,\Delta \nu)={S}_2(\u_n,\Delta \nu) + {F}_2(\u_n,\Delta \nu)
+{N}_2(\u_n,\Delta \nu)
\label{eq:vvc}
\end{equation}
where ${S}_2$, ${F}_2$ and ${N}_2$ respectively refer to the signal,
foreground and noise contributions to the visibility correlation. We
individually discuss the predictions for each of these components. 

\subsection{The HI signal}

\begin{figure}[t]
\begin{centering} 
\includegraphics[width=120mm,height=8.cm,angle=0.]{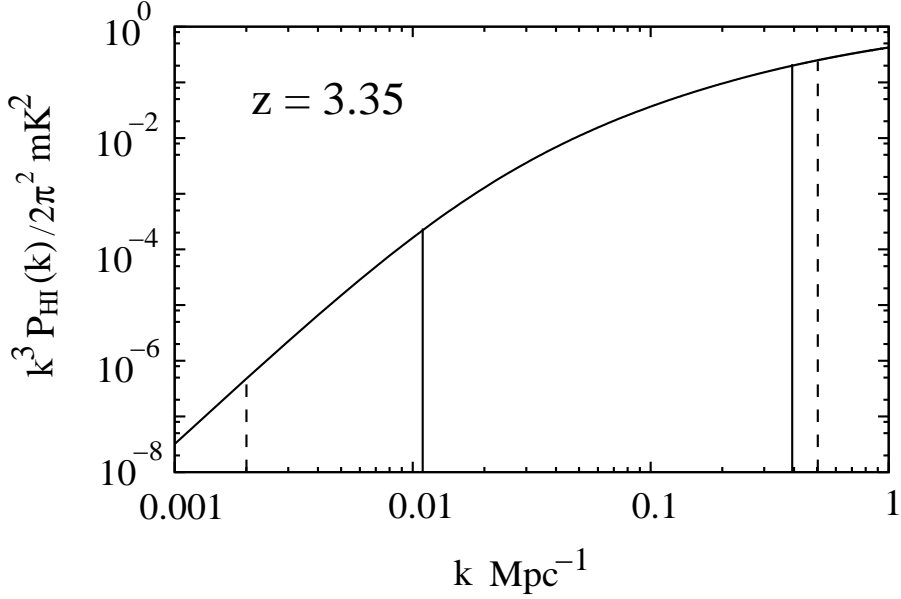}
\caption{ The solid curve shows the HI signal $k^3 \, P_{HI}(k)/ 2\,\pi^2$ where  
$P_{HI}(k) \equiv  P_{HI}(k,\mu=0)$ is the HI  power spectrum
  (eq. \ref{eq:pk}) at 
  $z=3.35$ which corresponds to $\nu_o=326.5 \, {\rm MHz}$. 
 The vertical  lines demarcate  the
$\kpr=\frac{2\,\pi\,\u}{\rn}$ range that  will be probed through
the HI signal at ORT. The solid and dashed vertical lines refer to
Phases I and II respectively. }  
\label{fig:pk.eps} 
 \end{centering}
\end{figure}

\begin{figure}
\begin{centering} 
\includegraphics[width=120mm,angle=0.]{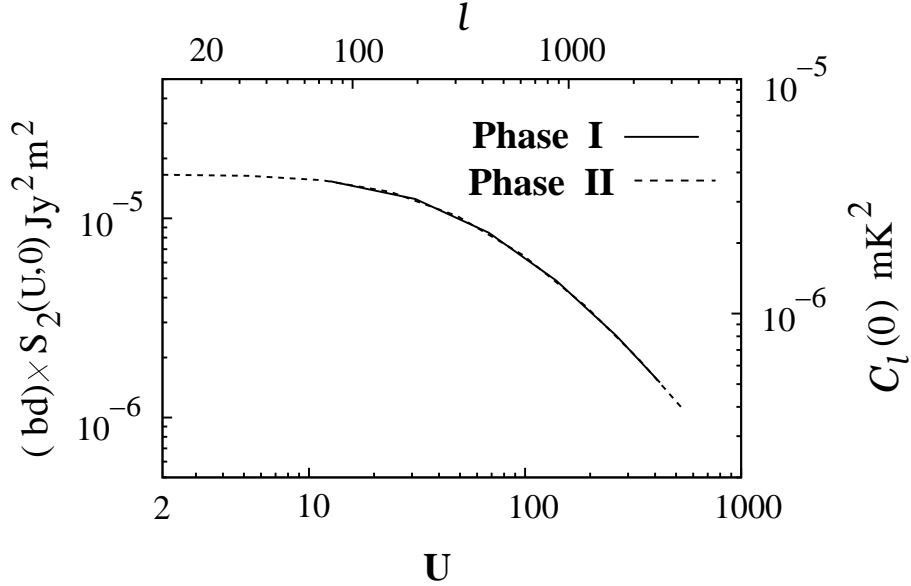}
\caption{ This shows $C_{\ell}(0)$ and also   $(b d) \times {S}_2(\u,0)$ 
which is expected to be independent of the size of the antenna
aperture.  We find that the values of $(b d) \times {S}_2(\u,0)$  predicted  
for  Phase I are nearly 
identical to those predicted for Phase II, and it is not possible to
distinguish between the two curves in the figure.  However, as shown
in the figure, the $U$ range  covered by Phase I is smaller.}
\label{fig:signal.eps} 
  \end{centering}
\end{figure}

\begin{figure*}
\begin{centering} 
\includegraphics[width=120mm,angle=0.]{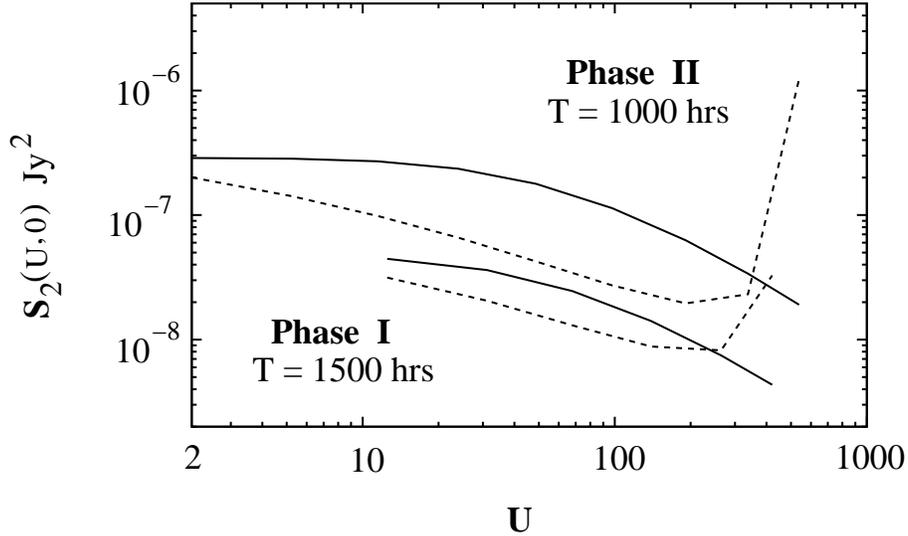}
  \caption{ This solid curves shows the expected HI signal
    ${S}_2(\u,\Delta \nu)$ for $\Delta \nu =0$ while the dashed
    curves show the $1-\sigma$ errors for the observation time
    indicated in the figure. The lower and upper sets of 
    curves correspond to Phases I and II respectively.} 
  \label{fig:signalnoise} 
  \end{centering}
\end{figure*}

The contribution ${S}_2(\u_n,\Delta \nu)$ to the visibility correlation
${V}_2(\u,\Delta \nu)$
from the HI signal  directly probes the three-dimensional (3D) power
spectrum $P_{\rm HI}({\bf k},z)$ of  the HI distribution in redshift
space (Bharadwaj \& Sethi 2001; Bharadwaj \& Ali 2005). It is
convenient here to use 
 \begin{equation}
C_{\ell}(\Delta \nu)= \frac{1}{\pi
  r_{\nu}^2}\, \int_{0}^{\infty} {\rm d} k_{\parallel}\,\cos
(k_{\parallel}\, r'_{\nu}\, \Delta \nu) \, P_{\rm HI}({\bf k}) 
\label{eq:hicl}
\end{equation}
to calculate the multi-frequency angular power spectrum for the HI
signal (Datta, Roy Choudhury \& Bharadwaj 2007), and use this in 
eq. (\ref{eq:c5}) to calculate $S_2(\u_n,\Delta \nu)$. 
Here $r_{\nu}$ is the comoving distance corresponding to $z=(1420
\, {\rm MHz}/\nu)-1$, $\rnp=\frac{d\rn}{d\nu}$ and the 3D wave vector
$\k$ has components $k_{\parallel}$ and $k_{\perp}=\ell/r$ which are
respectively parallel and perpendicular to the line of sight, and
$\mu=k_{\parallel}/k$ is the cosine of the angle between $\k$ and the
line of sight.  

We model $P_{\HI}(k,\mu) \equiv P_{\HI}(\k)$ assuming that the HI
traces the total matter 
distribution with a linear, scale-independent  bias parameter $b$. We
then have  
\begin{equation}
 P_{\HI}(k,\mu)=b^2\,\bar{x}^2_{\HI} \, {\bar{T}^2} \, \left[ 1+
   \beta\,  {\mu^2} \right]^2\,P(k) 
\label{eq:pk}
\end{equation}
where $P(k)$ is the matter power spectrum at the redshift $z$,
$\bar{x}_{\HI}$ is the mean 
hydrogen neutral  fraction and 
\begin{equation}
\bar{T}(z)=4.0 \, {\rm mK} \, (1+z)^2
\, \left(\frac{\Omega_b h^2}{0.024} \right) \, 
 \left(\frac{0.7}{h} \right) \, \left( \frac{H_0}{H(z)} \right) \,.
\end{equation}  
We  have used the  value $\bar{x}_{\HI} =2.45 \, 
\times \, 10^{-2}$ which corresponds to $\Omega_{gas}=10^{-3}$ (Noterdaeme et al. 2012; Zafar et al. 2013). 
  The term $\left[ 1+ \beta \mu^2 \right]^2$  arises due to  of  
the HI peculiar velocities (Bharadwaj, Nath \& Sethi
2001; Bharadwaj \& Ali 2004) and $\beta$ is the linear distortion
parameter. The semi-emperical estimate of the large-scale bias (b) of HI at redshift $z \sim \,3$ is 1.7 (Mar{\'{\i}}n et al. 2010). N-body simulations (Bagla, Khandai \& Datta  2010; Guha Sarkar et al. 2012) indicate that it is reasonably well justified to assume a constant HI bias $b=2$ at wave numbers $k \le 1 \, {\rm Mpc}^{-1}$, and we have used  this value for our entire analysis. The later result is consistant with Mar{\'{\i}}n et al. (2010).   

Figure \ref{fig:pk.eps}  shows $k^3 P_{HI}(k)/2{\pi}^2$ which quantifies
the magnitude of the expected HI signal.  In this figure we have fixed 
$\mu=0$ (eq. \ref{eq:pk}) which implies that the wave vector $\k$ is 
perpendicular to the line of sight.  The figure also shows  
the range of comoving wave numbers
$k$ where the  HI  power spectrum $P_{\rm HI}(k)$ will be probed by
the ORT. We see that the upper limit $\sim 0.5 \, {\rm   Mpc}^{-1}$ 
is  comparable    in both Phases I and II. The lower limit $\sim
0.002 \, {\rm   Mpc}^{-1}$, however, is considerably smaller for Phase
II in 
comparison to Phase I which is only sensitive to modes $k > 0.01 \, {\rm
  Mpc}^{-1}$.  We note that  Phases I and II are both sensitive to the BAO feature which has the first peak at
$k = 0.045 \, {\rm   Mpc}^{-1}$, and which has successive oscillations
whose amplitude decays within   $k = 0.3 \, {\rm  Mpc}^{-1}$ which is well
within the $k$ range that will be probed by ORT. It is planned to
investigate the possibility of detecting the BAO feature and 
constraining cosmological parameters in a separate, future study. 

Our subsequent discussion is in terms of the angular multipole $\ell$,
baseline $U$ and frequency separation $\Delta \nu$. Here we  briefly
discuss  how these quantities are related to the comoving wave numbers
$k_{\perp}$, $k_{\parallel}$ and $\Delta r_{\parallel}$ which is the
comoving distance 
interval along the line of sight. We have 
\begin{equation}
k_{\perp}=\frac{2 \pi U}{r_{\nu}} = \frac{\ell}{r_{\nu}}
\end{equation}
which relates $\ell$ and  $U$ to $k_{\perp}$ which is the wave
number perpendicular to the line of sight. Further,  we  have 
\begin{equation}
\Delta r_{\parallel} = r^{'}_{\nu} \, \Delta \nu = \frac{c \, \Delta
  \nu }{\nu_e a^2   H(a)} 
\end{equation}
which relates $\Delta \nu$ to $\Delta r_{\parallel}$. The comoving
wave number $k_{\parallel}$, which is parallel to the line of sight,
is the Fourier conjugate of $\Delta r_{\parallel}$. We have
$r_{\nu}=6.67 \, {\rm Gpc}$ and $ r^{'}_{\nu}=11.33 \, {\rm Mpc \,
  MHz^{-1}}$ at $\nu_0=326.5 \, {\rm MHz}$ for which the conversion
factors are summarized in Table \ref{tab:conv}. 

\begin{table}
\begin{center}
\caption{Conversion factors for $\nu_o=326.5 \, {\rm MHz}$.}
\label{tab:conv}
\vspace{.2in}
\begin{tabular}{|l|c|c|}
\hline
$k_{\perp}$  & $U$  & $\ell$ \\
\hline
$0.01 \, {\rm Mpc}^{-1}$ & $11$ & $67$ \\
\hline
\hline
$\Delta \nu$ & $\Delta r_{\parallel}$ & \\
\hline
$0.1 \, {\rm MHz}$ & $1.13 \, {\rm Mpc}$ & \\ 
\hline
\end{tabular}
\end{center}
\end{table}

We have used the HI power spectrum  $P_{HI}(\k)$ to calculate the
multi-frequency angular power spectrum $C_{\ell}(\Delta \nu)$ 
(eq. \ref{eq:hicl}). The HI signal is maximum when $\Delta \nu=0$, and 
Figure \ref{fig:signal.eps}  shows $C_{\ell}(0)$  as a function of
$\ell$. The respective $\ell$ range that will be probed by Phases I and II
is  also indicated in the figure. The value of $C_{\ell}(0)$ is around
$4 \times 10^{-6} \, {\rm mK}^2$ at the smallest $\ell$ values $(\ell
\lsim 150)$ where $C_{\ell}(0)$  is nearly constant independent of
$\ell$. Beyond this, the value of $C_{\ell}(0)$ decreases gradually,
and we have around $4 \times 10^{-7} \, {\rm mK}^2$ at $\ell \approx
3,300$ which is near the largest $\ell$ mode that will be probed. 

We now discuss   the visibility correlation HI signal
${S}_2(\u,\Delta \nu)$ predicted for the 
ORT (eq. \ref{eq:c5}). The signal is maximum for $\Delta \nu =0$, and
we first study   
${S}_2(\u,0)$ as a function of $U$  as shown in  Figure
\ref{fig:signalnoise}. 
For Phase II we see that  ${S}_2(\u,0)$ has a value $3 \times
10^{-7} \, {\rm   Jy}^2$  at the smallest baseline, and it is nearly
constant at small 
baselines $U \lsim 30$ beyond which it slowly falls to a value of around 
$2 \times 10^{-8} \, {\rm  Jy}^2$ at $U \approx 550$.
The $U$ range  is considerably smaller   for phase I where
${S}_2(\u,0)$ has a value $\sim 4 \times 10^{-8} \, {\rm Jy}^2$  at the
smallest baseline $(U \approx 10$) and falls to $\sim 5 \times 10^{-9}
\, {\rm Jy}^2$  at $U \approx 420$.

The visibility correlation  ${V}_2(U,\Delta \nu)$ depends on the size
of the antenna aperture through the factor $|\tilde{A}(\u, \nu)|^2$
which appears in  eq. (\ref{eq:c5}). The amplitude (eq. \ref{eq:a2}) 
scales as   $|\tilde{A}(\u, \nu)|^2 \propto (b d)^{-2}$  whereas 
the region in $\u$ space where this function has support scales as
$\propto (b d)$.  As a consequence  it follows that we expect the
scaling $ {V}_2(\u, \Delta \nu) \propto (b d)^{-1}$, and we expect 
$(b d) \times  {V}_2(\u, \Delta \nu)$ to be independent of the size
of the antenna aperture. This is also apparent from eq. (\ref{eq:c7})
which relates $(b d) \times  {V}_2(\u, \Delta \nu)$ to
$C_{\ell}(\Delta \nu)$, and which is expected to hold if the  baselines $U$ is
sufficiently large.   Figure \ref{fig:signal.eps}  shows $(b d) \times
{S}_2(\u,0)$ as a   function of $U$. We find that the predictions for
Phases I are   nearly identical to those for Phase II, and it is not
possible to  distinguish between the two curves in the figure.
Based on this we conclude that  it is adequate to use eq. (\ref{eq:c7}) 
to relate the  visibility correlation $S_2(U,0)$ to the
multi-frequency angular power spectrum $C_{\ell}(0)$ of the HI
signal for the entire baselines baseline range at ORT. Further,  we also expect this   to
hold for other values of  $\Delta \nu$, as well as for the foreground
${F}_2(U,\Delta \nu)$.

We now quantify the $\Delta \nu$ dependence of the HI signal 
${S}_2(\u,\Delta \nu)$. We use the frequency decorrelation function 
$\kappa_{\u}(\Delta \nu)$  (Datta, Roy Choudhury \& Bharadwaj 2007)
which is defined as  
\begin{equation}
\kappa_{\u}(\Delta \nu)=\frac{S_2(U,\Delta \nu)}{S_2(U, 0)}\,.
\label{eq:ka}
\end{equation}

\begin{figure}[t]
\begin{centering} 
\includegraphics[width=120mm,angle=0.]{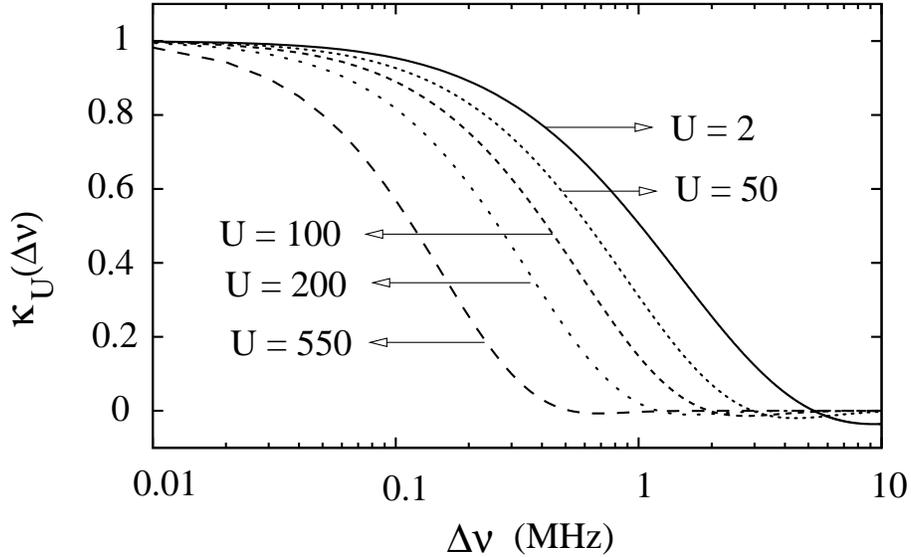}
\caption{ This figure shows the predicted frequency decorrelation
  function  $\kappa_{\u}(\Delta \nu)$ as a function of $\Delta \nu$
  at five different $\u$ values. The signal
  decorrelates  more sharply for higher value of $\u$.}  
 \label{fig:kappa}
  \end{centering}
\end{figure}

This function quantifies  how quickly the HI signal decorrelates
as we increase the frequency separation $\Delta \nu$. The signal is
perfectly correlated at $\Delta \nu=0$ where we have
$\kappa_{U}(0)=1$, and the correlation falls  ($\kappa_{U}(\Delta
\nu) < 1$) as $\Delta \nu$ is increased. Figure \ref{fig:kappa} shows
the variation of   $\kappa_{U}(\Delta \nu) $ as a function of
$\Delta \nu$ for different values of $U$. Here we have assumed the 
statistics of the  HI signal is stationary across frequency, and thereby 
only  depends on $\Delta \nu$. The predictions are the same
for Phases I and II, and we do not show separate curves for the two
phases.  

\begin{figure}[t]
\begin{centering} 
\includegraphics[width=120mm,angle=0.]{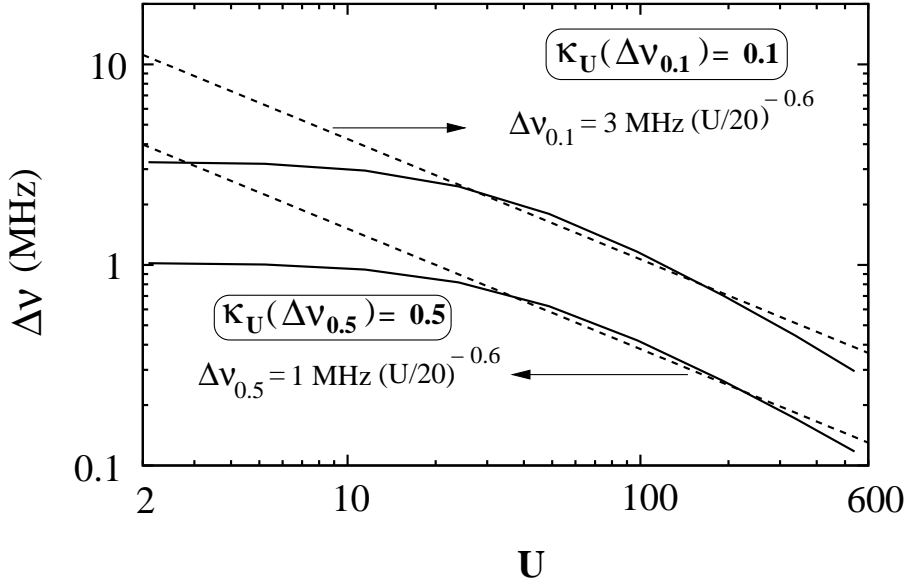}
\caption{This figure shows how $\Delta \nu$ are varying with functions
  of $U$ for a given value of $\kappa_{\u}(\Delta \nu)$. The upper and
  lower solid curves correspond to $\kappa_{U}(\Delta\nu_{0.1}) =0.1$
  and $\kappa_{U}(\Delta\nu_{0.5})=0.5$ respectively. The definition
  of $\Delta\nu_{0.1}$ and $\Delta\nu_{0.5}$ are given in the
  text. The dotted lines are are power law fitting for
  $\Delta\nu_{0.1}$ and $\Delta\nu_{0.5}$ to quantify how rapidly the
  signal decorrelates at different $\u$ values.}
 \label{fig:kappa.5}
  \end{centering}
\end{figure}

We see that at the smallest baseline $U \approx 2$ the value of
$\kappa_{U}(\Delta \nu)$  decreases slowly as $\Delta \nu$ is
increased. We have $\kappa_U = 0.5$ at $\Delta \nu \approx 1 \,
{\rm MHz}$, beyond which the value of $\kappa$  falls further. The
value of $\kappa_U$ crosses zero at around $\Delta \nu \approx 4 \, {\rm
  MHz}$, and  $\kappa_U$ becomes negative beyond this. The
decorrelation function $\kappa_{U}(\Delta \nu)$  shows a similar
$\Delta \nu$ dependence  at larger baselines, with the difference that
we have a steeper $\Delta 
\nu$ dependence  at larger baselines. For $U = 200$, we see that
$\kappa_U =0.5$  at $\Delta \nu \approx 0.2 \, {\rm MHz}$  and
$\kappa_U$ crosses zero 
well before $\Delta \nu = 1 \, {\rm MHz}$. We also see that the value
of $\kappa_U$ oscillates round zero for large values of $\Delta \nu$.  
We have defined  $\Delta _{0.5}$ and $\Delta _{0.1}$  as the values of
the frequency separation $\Delta \nu$ where the decorrelation
falls to $0.5$ and $0.1$ respectively {\it ie.}  $\kappa_{U}(\Delta
\nu_{0.5})   =0.5$, etc. Figure \ref{fig:kappa.5} shows $\Delta
\nu_{0.5}$  and $\Delta\nu_{0.1}$ as  functions of  
$U$. We use  $\Delta\nu_{0.5}$ to compare and quantify how rapidly the
signal decorrelates at different values of $U$. We see that $\Delta 
\nu_{0.5}$ has a  nearly constant value $\approx 1 \, {\rm MHz}$ for 
$U \le 30$, and it declines as $U^{-0.6}$ for larger baselines in the
range of our interest. The value of $\Delta \nu_{0.1}$ gives an
estimate of the  frequency  separation across which the HI signal is
correlated, and  the bulk of the HI signal is contained within
$\Delta \nu \le \Delta \nu_{0.1}$. We see that $\Delta \nu_{0.1}$ has
a nearly constant value  $\approx 3 \, {\rm MHz}$ at the small
baselines ($U \le 30$), and we have $\Delta \nu_{0.1} \approx 3 \times
\Delta \nu_{0.5} $ for the entire baseline range of our interest. 

Figures \ref{fig:kappa} and  \ref{fig:kappa.5} provides an estimate of 
two of the system parameters, namely the total frequency bandwidth $B$
and the frequency channel width $\Delta \nu_c$ 
desirable for quantifying  the HI signal. We see that the HI signal
remains correlated to frequency separations as large as $3 \, {\rm
  MHZ}$ at  the small baseline. It is thus desirable
to have a bandwidth $B$ larger than this, which is well within the
specifications of both Phases I and II (Table~\ref{tab:array}).
Considering the channel width, we see that $\Delta \nu_c = 10 \, {\rm
  kHz}$ 
or $20 \, {\rm kHz}$  is small enough to adequately  quantify the
decorrelation of the HI signal at even the largest baseline of our
interest. A larger channel width of  $\Delta \nu_c = 200 \, {\rm kHz}$
will be adequate for baselines $U \le 200$, however some of the signal
would be missed out at the large baselines ($U \sim  500$). 

\subsection{Noise and error estimates}
We first consider the noise contribution ${N}_2(\u_n,\Delta \nu)$ to
the visibility correlation (eq. \ref{eq:vvc}).   The noise in the 
visibilities measured at different antenna pairs is uncorrelated. 
As noted earlier, the ORT has a high degree of redundancy and there
are many independent antenna pairs corresponding to the same
baseline. Further, an observation spanning a total observing time of
$t_{\rm obs}$  provides $t_{\rm obs}/\Delta t$ different measurements
of each visibility. The noise in the visibilities measured at two
different time instants is uncorrelated. It is possible to 
avoid the  noise contribution  ${N}_2(\u_n,\Delta \nu)$ in the
visibility correlation ${V}_2(\u_n,\Delta \nu)$  by 
correlating only the visibility measurements where the noise is
uncorrelated   (eg. Begum et al. 2006, Ali et al. 2008).  For a
fixed baseline $\u$ we only correlate the visibilities measured by
different antenna pairs or the visibilities measured at different time
instants. We therefore do not include  the noise  
contribution to the visibility correlation in the
subsequent analysis, and use 
\begin{equation}
{V}_2(\u_n,\Delta \nu)={S}_2(\u_n,\Delta \nu) + {F}_2(\u_n,\Delta
\nu) \,.
\label{eq:vvc1}
\end{equation}
The noise however contributes to the uncertainty in the visibility
correlation $\sqrt{(\Delta {{V}_2})^2}$ which we
discuss below. 

We now calculate the expected statistical fluctuations (errors) or
uncertainty  in the estimated visibility correlation
${\V}_{2}(U,\Delta \nu)$. It is assumed that the foregrounds have been
completely removed 
from the visibilities whereby  the residuals  after foreground
subtraction  contain only the HI signal and system noise. Therefore
the total error in the residual visibility correlation  has two parts 
arising from  the cosmic variance and the system noise
respectively. The expected 
uncertainty or statistical fluctuations in 
the visibility correlation   is
\begin{equation}
\sqrt{ (\Delta {{V}_2})^2 } = \sqrt{(\Delta
  {S}_2)^2 + (\Delta {N}_2)^2 }\,,
\label{eq:tn}
\end{equation}
where $(\Delta{S}_2)^2$ and $(\Delta {N}_2)^2$ are the 
cosmic variance  and the system noise contributions  respectively. 

 We have $(\Delta {N}_2)^2 =(2 \sigma^2)^2$ and $(\Delta {S}_2)^2 =
({S}_2)^2$ for a single estimate of the visibility correlation.
The system noise contribution reduces to $(\Delta {N}_2)^2 =(2
\sigma^2 \Delta t/t_{\rm obs})^2$ if we combine the measurements at
different   time instants. 
The  redundant baselines provide many estimates of the  visibility correlation
${V}_2(\u,\Delta \nu)$ at the same  $\u$. Each estimate has an independent 
system noise contribution, the  signal however is the the same . 
 We also bin the data by combining the estimates  of ${V}_2(\u,\Delta \nu)$  at
 the different  $\u$ values   within a finite bin width of our choice.  The 
different baselines  $\u$ provide  independent estimates of both the 
signal and the system noise.  We use $N_P$ and  $N_E$ respectively to denote 
the  number of  independent estimates of the system noise and the signal in 
 ${V}_2(U,\Delta \nu)$  for each  bin.  

The frequency bandwidth $B$ also provides several independent
estimates of the visibility correlation. The value of $\Delta
\nu_{0.5}$ provides an estimate of the frequency separation over which
the HI signal  remains correlated. For the purpose of the estimates
presented here we assume a channel width of  $\Delta \nu_c=\Delta \nu_{0.5}$ 
in  eq. (\ref{eq:rms}) for $\sigma^2$ and  also 
assume that the frequency bandwidth $B$ provides us
with $B/(\Delta \nu_{0.5})$ independent estimates of the visibility
correlation. Combining all the effects discussed above we have 
\begin{equation}
(\Delta {N}_2)^2 = \left( \frac{2 \sigma^2 \,  \Delta t}{t_{\rm obs}} \right)^2
\frac{\Delta \nu_{0.5}}{N_P B}
\label{eq:n4}
\end{equation}
and 
\begin{equation}
(\Delta {S}_2)^2 = \frac{({S}_2)^2 \Delta \nu_{0.5}}{N_E B}
\label{eq:cvar}
\end{equation}
which we use in eq.(\ref{eq:tn}) to calculate the error estimates
$\sqrt{(\Delta {V}_2)^2}$ for ${S}_2(U,0)$   and also 
the signal to noise ratio SNR$={S}_2(U,0)/ \sqrt{(\Delta {V}_2)^2}$. 
We have calculated  $N_P$ and $N_E$   by dividing  the
baseline range $U_{min}$ to $U_{max}$  into $6$ and $9$ logarithmic
bins for Phases I and II respectively.

\begin{figure*}
 \vskip.2cm \centerline{{ \includegraphics[width=7.cm,height=5.cm
     ]{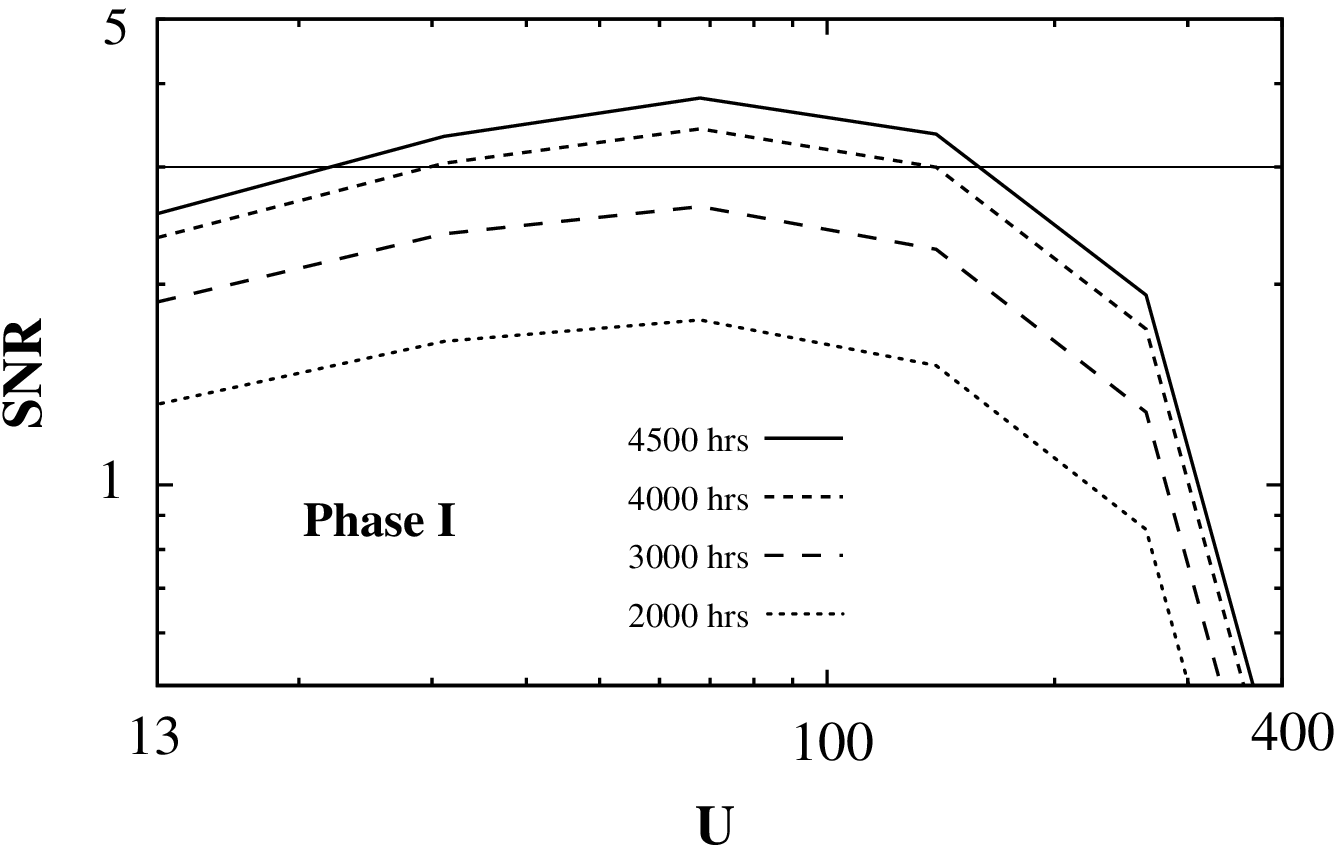}} \hskip1.cm {
     \includegraphics[width=7.cm,height=5.cm
     ]{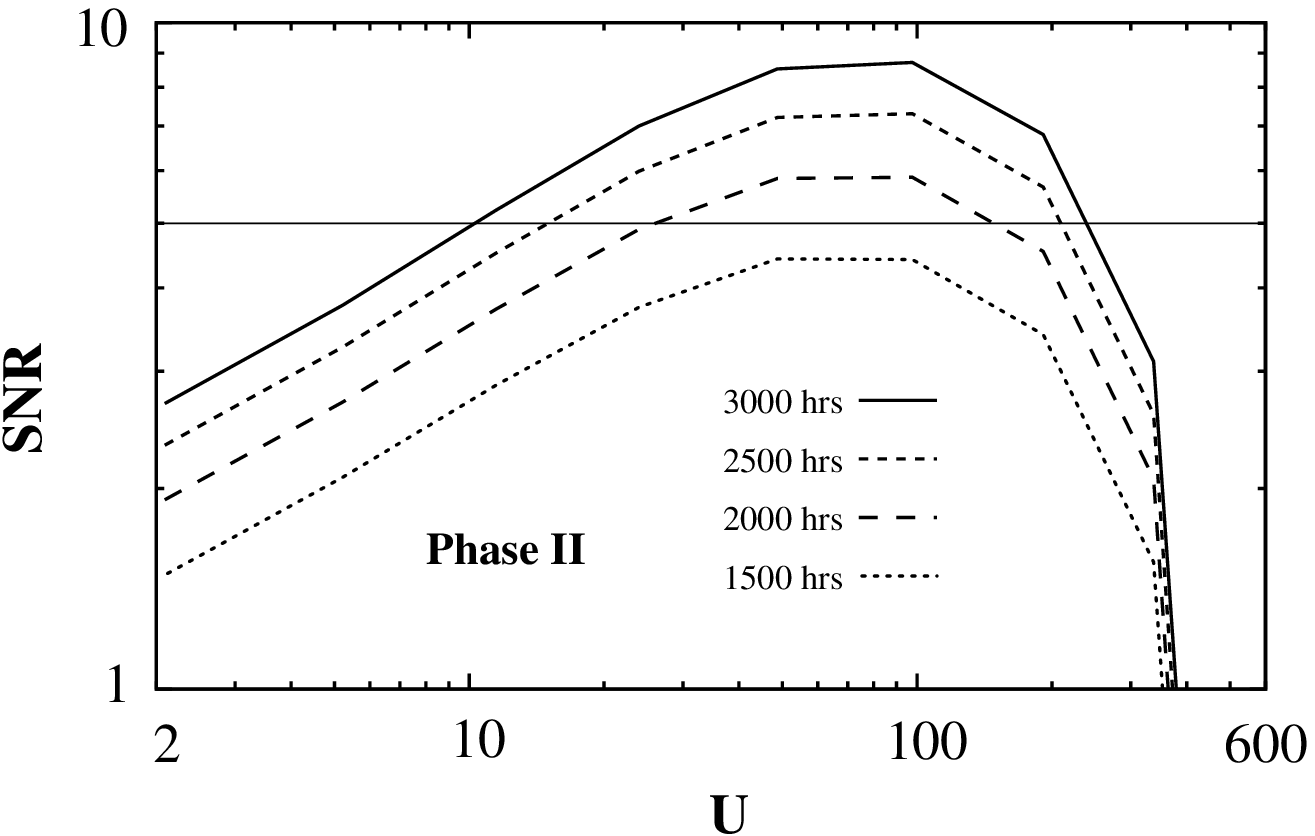}} }
 
\caption{This figure shows the signal to noise ratio (SNR) as function
  of baseline $\u$ for different integration times are indicated. The
  horizontal line is for $\rm SNR=5$ (right panel.) and $\rm
  SNR=3$ (left panel.) }
\label{fig:signaltonoiseratio_u} 
\end{figure*}

The uncertainty $\sqrt{(\Delta {V}_2)^2}$ shown in Figure 
\ref{fig:signalnoise} is dominated by  the system noise over the
entire baseline range.   The SNR, shown in Figure
\ref{fig:signaltonoiseratio_u}, peaks at $U \sim 100$ which
corresponds to $\sim 30^{'}$.  This peak feature is particularly
prominent for Phase II which also has a higher SNR compared to Phase
I. We see that for Phase I, a $3 \sigma$ detection is possible  at $U
\sim 100$ with $\sim 4,000 \, {\rm hrs}$ of observation. The SNR
scales as $\propto t_{\rm obs}$, and a $5 \sigma$ detection is 
possible with $\sim 5,700 \, {\rm hrs}$ of observation with Phase I.

The signal and  noise in the individual visibilities are both larger  for Phase II in
comparison to Phase I (Table~\ref{tab:array}). The noise contribution to a single visibility 
scales as $\sigma \propto (b d)^{-1}$ (eq. \ref{eq:rms})
whereas the signal scales as  $\sqrt{{V}_2}\propto (b d)^{-0.5}$  (eq. \ref{eq:c7}), and 
 a single visibility has a lower  SNR in  Phase II as compared to Phase I.   
 However, there is a substantial increase in the number of baselines  for Phase II 
which more than compensates for this,  and  we find that the binned visibility correlation 
has a considerably higher   SNR  for Phase II in  comparison to Phase I (Figure
\ref{fig:signaltonoiseratio_u}).  We see 
that a $3 \sigma$ detection is possible with $\sim 1,000 \, {\rm hrs}$ of observation 
with Phase II. A $5 \sigma$ detection is possible in the baseline range $40 \le U \le 100$ with  
$\sim 2,000 \, {\rm hrs}$ of observation. It is possible to detect the HI signal at a
significance greater than $5 \sigma$  in the  baseline range $10 \le U \le 200$ 
with  $\sim 3,000 \, {\rm hrs}$ of observation.

\subsection{Foregrounds}
We refer to  the radiation  from different astrophysical sources  other than
the cosmological HI signal  collectively   as foregrounds.
Foregrounds  include extragalactic point sources,
diffuse synchrotron radiation from our Galaxy and low redshift galaxy
clusters; free-free emission from our Galaxy (GFF) and external
galaxies (EGFF). Extra-galactic point sources and the diffuse
synchrotron radiation from our  Galaxy  largely dominate  the
foreground radiation at $ 326.5\, \rm{MHz}$. The free-free
emissions from our Galaxy and external galaxies make much smaller
contributions though each of these is individually larger than the HI
signal. All the foreground components mentioned earlier 
are continuum sources. It is well accepted that the frequency dependence of 
the various continuum foreground  components can be modelled  by power laws, 
and we  model the multi-frequency angular 
power spectrum for each foreground component as
\begin{equation}
C_{\ell}(\nu_1,\nu_2)=A\left(\frac{1000}{\ell}\right)^{\gamma}
\left(\frac{\nu_f}{\nu_1}\right)^{\alpha} 
\left(\frac{\nu_f}{\nu_2}\right)^{\alpha}
\label{eq:fg}
\end{equation}
where $A$ is the amplitude in
$\rm mK^2$, and   $\gamma$  and $\alpha$ are the power law indices  for the $\ell$ 
and the $\nu$ dependence respectively. In the present analysis we are interested in a
 situation where $\nu_2=\nu_1 + \Delta \nu$ with $\Delta \nu \ll \nu_1$, and we have  
\begin{equation}
C_{\ell}(\Delta \nu) \equiv C_{\ell}(\nu_1,\nu_1 + \Delta \nu) \approx A\left(\frac{1000}{\ell}\right)^{\gamma}
\left(\frac{\nu_f}{\nu_1}\right)^{2 \alpha} \left(1 - \frac{\alpha \, \Delta \nu }{\nu_1}
\right)
\label{eq:fga}
\end{equation}
which varies slowly with $\Delta \nu$. For the foregrounds, we expect 
$C_{\ell}(\Delta \nu)$ to fall  by less than $10 \%$ if $\Delta \nu$ is varied
 from $0$ to $3 \, {\rm MHz}$, in contrast to the $\sim 90 \%$ decline 
predicted for the 
\HI  signal (Figure \ref{fig:kappa.5}). 
The  frequency spectral index $\alpha$ is expected to have a scatter $\Delta \alpha$ in 
the range $0.1 -0.5$ for the different foreground components in  different directions  
causing  less than  $2 \%$  additional
deviation in the frequency band of our interest. We  only consider the mean spectral indices
for the purpose of the foreground predictions presented here. In a nutshell, the 
 $\Delta \nu$  dependence  of $C_{\ell}(\Delta \nu)$ is  markedly different for the  
foregrounds as compared to the \HI signal. and we hope to use this to
 separate  the foregrounds from the \HI signal. 

There are, at present, no observational constraints on the  $\Delta \nu$ 
dependence  of $C_{\ell}(\Delta \nu)$ for any of the foreground components
at the angular scales and frequencies of our interest.  
We do not attempt to make any model predictions for the $\Delta \nu$ 
dependence  beyond assuming that $C_{\ell}(\Delta \nu)$  varies slowly 
with $\Delta \nu$ across the frequency separations  of our interest. 
For the present work we focus on  $C_{\ell}(0)$ which we have modelled as
\begin{equation}
C_{\ell}(0) =A\left(\frac{1000}{\ell}\right)^{\gamma} \, ,
\label{eq:clfg}
\end{equation}
and we assume the that the $\Delta \nu$ dependence is very slow 
whereby  the foregrounds can  be separated from the \HI signal. 
In the
subsequent discussion we focus on model predictions for $A$ and $\gamma$
which are tabulated in Table \ref{tab:parm} for the different
foreground components. The values of $A$, whenever used
in this paper, have all been scaled $(A \propto \nu^{-2 \alpha})$ 
to  the nominal frequency of  $\nu_o=326.5 \, {\rm  MHz}$  using the 
$\alpha$ values   tabulated in Table \ref{tab:parm}.

Extra-galactic point sources are expected to dominate the $326.5 \, \rm MHz$
sky at the angular scales of our interest. 
 The contribution from extragalactic point 
sources is mostly due to the emission from normal galaxies, radio
galaxies, star forming galaxies and  active galactic nuclei (Santos et
al. 2005). Predictions of the point source contribution are based on the measured 
source count function and the angular correlation function.

There are different radio surveys that have been  conducted at various frequencies 
ranging   from $151 \,{\rm MHz}$ to $8.5 \,{\rm GHz}$,  and these have a
wide range of angular resolutions ranging  from $1^{''}$ to $5^{'}$ (eg. Singal et
al. 2010, and  references therein). There is a clear consistency among the 
differential source count functions ($ \frac{dN}{dS} \propto
S^{-\epsilon}$) at $1.4 \,{\rm GHz}$ for sources with flux $S > 1 \,{\rm mJy}$. 
The source counts are poorly constrained at
$S \,< 1 \,{\rm mJy}$.  Based on the various radio observations
(Singal et al. 2010), we have identified  four different regimes  for the  $1.4 \,{\rm
  GHz}$ source counts (a.)$\, \gsim 1\,{\rm Jy}$  which are the brightest
sources in the catalogs. These   are relatively nearby objects and they 
have a  steep, Euclidean  source count with $\epsilon \sim 2.5 $;
(b.) $1\, {\rm mJy}$ - $ 1\, {\rm Jy}$ where the  observed
differential source counts decline more gradually  with $ \,\epsilon \sim 1.7
$ which  is caused by redshift effects; (c.) $ 15 \, \mu {\rm Jy}$ - $\, 1\,
{\rm mJy}$ where the source counts are again steeper  with $\epsilon > 2$  which is 
closer to Euclidean,  and there is considerable scatter from field to field; 
 and (d.) $\lsim\, 15 \,\mu {\rm Jy}$,  the source counts must eventually flatten 
($\epsilon < 2$) at  low $S$ 
to avoid an infinite integrated flux. The 
cut-off  lower flux where the power law index $\epsilon$
falls  below $2$ is not well established, and  deeper radio observations are
 required. 

 If we extrapolate the $ 1.4\, \rm GHz$ source counts to $ 326.5\,
\rm MHz$,  the power law index $\epsilon$  remains unchanged, but the flux
range and the constant of proportionality 
change. This change depends upon the value of the frequency spectral index 
used to   extrapolate from $ 1.4\, \rm GHz$ to  $ 326.5\, \rm MHz$.

The first turnover or flattening in the  $ 1.4\, \rm GHz$ 
differential source count has been reported 
at $\sim \,1\,\rm mJy$    (Condon 1989; Rowan-Robinson
et al. 1993; Hopkins et al. 1998; Richards 2000; Hopkins et al. 2003;
Seymour, McHardy \& Gunn 2004; Jarvis \& Rawlings 2004; Huynh et
al. 2005; Simpson et al. 2006; Owen \& Morrison 2008).  
The flatenning  is attributed to the emergence of a new
population of radio sources (star-forming galaxies and low-luminosity
AGN) below $\sim \,1\,\rm mJy$.  
 The turnover flux of $1 \, {\rm  mJy}$ at $ 1.4\, \rm GHz$ is equivalent to
$\sim 3 \,\rm mJy$ at $325\, {\rm MHz}$ assuming a spectral index of
0.7. This is consistent with $325\, {\rm MHz}$ GMRT observations
(Sirothia et al. 2009). This features has also been observed at  $S \sim \,1.9\,\rm mJy$ 
in   $ 610\, \rm MHz$ GMRT observations (Garn et al. 2007; Bondi et al. 2007; Garn et
al. 2008). 

We have modelled the $326.5\, {\rm MHz}$  source count function
using a  double power law 
\begin{eqnarray} 
  \frac{dN}{dS}= \left \{\begin{array}{ll}
      \frac{4000}{Jy \cdot Sr}\cdot\,\left(\frac{S}{1
          Jy}\right)^{-1.64}\,\,\,\, {\rm for}\,&
      \hbox{$3 \, {\rm mJy}
        \le \,S \le 3 \, {\rm Jy}$ } \\ \frac{134}{Jy
        \cdot Sr}\cdot\,\left(\frac{S}{1 Jy}  
      \right)^{-2.24}\,\,\,\, {\rm for} & \hbox{$
        {10 \, \mu \rm Jy}
        \le \,S \le 3 \, {\rm mJy} \,.
        $}\\
    \end{array}\right.\, 
  \label{eq:sc}
\end{eqnarray} 
Here we have fitted  the $325 \,{\rm MHz}$ differential source count measured by 
Sirothia et al. (2009) to obtain the power law for $S \ge 3 \, {\rm mJy}$.  This measurement, 
which  is the deepest till date at this frequency, does not  adequately cover sources fainter 
than  $ 3 \, {\rm mJy}$.   For the sources  below $ 3 \,\rm mJy$, we fit $1.4 \,
  {\rm GHz}$ source counts from extremely deep VLA observations (Hubble
  Deep Field-North; Briggs \& Ivison 2006) in the flux range
  $30 \,\rm \mu Jy$ to $\,1153\,\rm \mu Jy$ by a single power law with
  slope $\epsilon=2.24$. We scale this  to $326.5
  \,{\rm MHz}$ using an average spectral index of 0.7 (Jackson 2005;
  Randall et al. 2012).  As mentioned earlier, the lower cut-off below which 
the source count flattens is not well determined. For the purpose of this paper, 
we  assume that the power law behaviour $S^{-2.24}$ holds for
 $S \ge 10\, \mu {\rm Jy}$, and the source count is flatter than $S^{-2}$
(we use $\epsilon =1.5$, Condon et al. 2012) for sources fainter than this. The choice of 
$10\, \mu {\rm Jy}$ is motivated by the fact that the  total contribution from 
sources with flux $S \le 10\,\mu {\rm Jy}$ to each pixel in the  sky
converges to $\sim 10\, \mu {\rm Jy}$ for a pixel size of 
$\sim 2^{'}$ which is comparable to the N-S resolution of the ORT. 

Point  sources make  two distinct contributions to the angular power
spectrum, the first being the   Poisson fluctuation  due to the discrete
nature of the sources and the second arising  from  the  clustering
of the sources.  
The Poisson contribution, which is independent of $\ell$, is
calculated using 
\begin{eqnarray}
C_{\ell}(0) =  \left(\frac{\del B}{\del   T}\right)^{-2} \left[ 
\int^{s_c}_{0}
 S^2 \frac{dN}{dS}\, dS \right]
\label{eq:ps}
\end{eqnarray}
where $S_{c}$ ($\le 3\, {\rm Jy}$) is the cut-off flux,    all point sources 
brighter than this  are assumed to have been  identified and subtracted out 
from the data. The Poisson contribution is dominated by the brightest sources 
$(S \sim S_c)$, and the $10\, \mu {\rm Jy}$ lower cut-off  does not make a
significant contribution to the amplitude $A$ listed in Table~\ref{tab:parm}.

\begin{table}
\caption{Values of the parameters used for characterizing different
  foreground contributions at $ 326.5 \,\rm MHz$.}
\vspace{.2in}
\label{tab:parm}
\begin{tabular}{|p{4cm}|p{5cm}|p{1cm}|p{1cm}|}
\hline \hline Foregrounds&$\,\,\,\,\,\,\,\,A({\rm
  mk^2})$&$\,\,\,\,{\alpha}$&$\,\,\,\,{\gamma}$\\ \hline Point source& $\,\,288
\left(\frac{S_{c}}{\rm Jy}\right)^{1.36}+0.01 $&$2.7$&$0$\\ (Poisson)  & & & \\ 
\hline Point source & $\,\,453 \left(\frac{S_{c}}{\rm Jy}\right)^{0.72}$ - $112 \left(\frac{S_{c}}{\rm Jy}\right)^{0.36}$&$2.7$&$0.9$\\ (Clustered)& +161 & & \\
 \hline Galactic synchrotron&$\,\,\,\,\,10.2
$&$2.52$&$2.34$\\ \hline Galactic
free-free&$\,\,\,\,1.7\times10^{-3}$&$2.15$&$3.0$\\ \hline Extra
Galactic free-free&$\,\,\,\,2.3\times10^{-4}$&$2.1$&$1.0$\\ \hline
HI signal & $1.1 \, \times \, 10^{-6}$ & - & - \\ \hline
\end{tabular}
\end{table}

The analysis of large samples of nearby radio-galaxies has shown that
the point sources are clustered. Cress et al. (1996) have measured the
angular two point correlation function at $1.4 \, \rm GHz$ (FIRST
Radio Survey, Becker et al. 1995)  across angular scales of
$1.2^{'}\, \rm to\, 2^{o}$, equivalent to a $\u$ range of
$14<\u<1430$. The measured two
point correlation function can be well fitted with a single power law
\begin{equation}
w({\theta})=(\theta/\theta_{0})^{-\beta}
\label{eq:omegatheta}
\end{equation}
 where  $\beta = 1.1$ and $\theta_{0} = 17.4'$. They have also reported
 that  the double and multi-component  sources tend to have a larger clustering amplitude 
than  the whole sample on small scales ($\le 12^{'}$). Further, 
the sources with flux  below $2\,\rm mJy$ have a  shallower slope 
($\beta \sim 0.97$).

 For the purpose of this paper we have modeled the angular power spectrum 
due to the  clustering  of point  sources  as
\begin{eqnarray}C_{\ell}(0) =  \left(\frac{\del B}{\del   T}\right)^{-2} \left[ 
\int^{s_c}_{0} S \frac{dN}{dS}\, dS \right] w_{\ell}\,\,
\label{eq:clus}
\end{eqnarray}
where $ w_{\ell} \propto {\ell}^{\beta -2}$ is  the Legendre
transform of $w(\theta)$. In this case, 
the amplitude $A$  listed in Table~\ref{tab:parm} is sensitive to both 
the upper cut-off $S_c$ and the lower cut-off of   ${10 \, \mu \rm Jy}$. 
However, in reality the  faint sources have a weaker clustering as compared to the single 
 power law adopted here, and  we do not expect  a  very
 significant dependence 
on the lower cut-off  of  ${10 \, \mu \rm Jy}$.

The Galactic diffuse synchrotron radiation is believed to be
produced by cosmic ray electrons propagating in the magnetic field of
the Galaxy (Ginzburg \& Syrovatskii 1969; Rybicki \& Lightman
1979). La Porta et al. (2008) have determined  the angular power spectra
of the Galactic synchrotron emission at angular scales greater 
than $0.5^{\circ}$  using total intensity all sky maps at $ 408 \,\rm MHz $
(Haslam et al. 1982) and $1.42 \, \rm GHz$ (Reich 1982, Reich \& Reich
1986; Reich, Testori \& Reich 2001). They find that the 
 angular power spectrum of synchrotron emission is well described 
by a power law   (eq. \ref{eq:clfg}) where the value 
of $\gamma$ varies  in the range $2.6$ to $3.0$ depending on the 
galactic latitude. Further, they have analyzed the frequency dependence 
to find $A \propto \nu^{-2 \alpha}$ with $\alpha$ varying in the range 
$2.9$ to $3.2$. 

The  angular power spectrum of the Galactic synchrotron radiation 
has been recently measured at  angular scales less than $0.5^{\circ}$
in three different   $150 \, \rm MHz$ observations. 
Bernardi et al. (2009) have estimated  $ \gamma=2.2 $ and
$A=253 \, {\rm mK}^2$ using WSRT  observations in  a field 
with Galactic latitude $b=8^{\circ}$. 
 Ghosh et al. (2012) have estimated  $\gamma = 2.34$ and 
$A=513 \, {\rm mK}^2$ using GMRT  observations in  a field 
with Galactic latitude $b=14^{\circ}$. Iacobelli et al. (2013) 
have  estimated  $\gamma = 1.84$ using LOFAR  observations in 
the same field as Bernardi et al. (2009). 
The mean spectral index of the synchrotron emission at high Galactic 
latitude has been recently  constrained to be $\alpha =2.52$ in
the $150 - 408 \, \rm MHz$ frequency range (Rogers \& Bowman 2008) using 
single dish observations. For the purpose of this paper we
have used $A$ and $\gamma$ from  Ghosh et al. (2012) and extrapolated
this to $326.5 \, {\rm MHz}$ using $\alpha$ from
Rogers \& Bowman (2008).  The parameters for the 
Galactic and Extra-Galactic free-free emission have been
extrapolated from $130 \, \rm MHz$ (Santos et al. 2005), and are listed
in Table \ref{tab:parm}.  For comparison, the value of $C_{\ell}(0)$ at $\ell=1,000$ for the 
\HI signal is also shown in  Table \ref{tab:parm}. 

\begin{figure}
\begin{centering} 
\includegraphics[width=120mm,angle=0.]{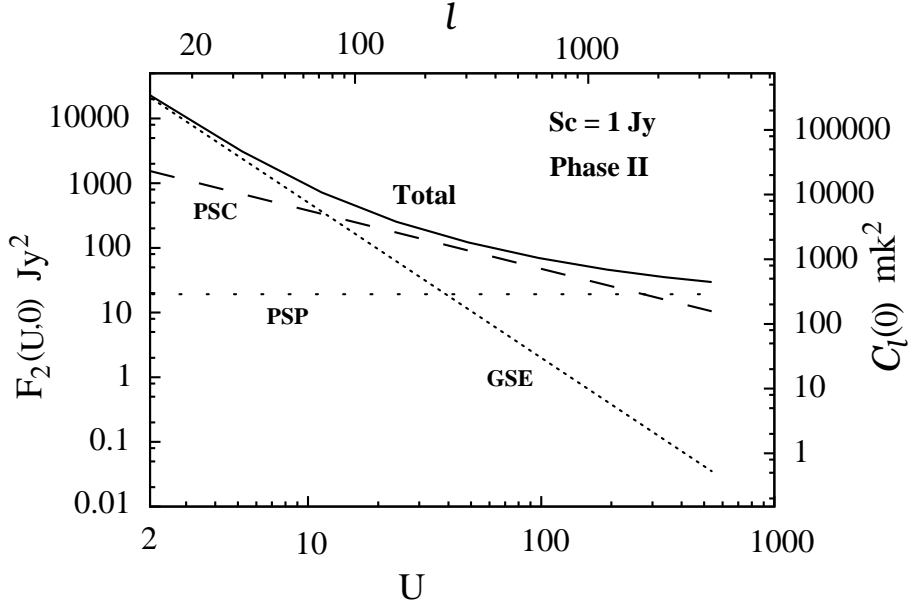}
\caption{This  shows the foreground model predictions for Phase II under the assumption that the 
brightest source in the field has a flux of  $S_c= 1 \, {\rm Jy}$. Th left and 
bottom axes respectively show   ${F}_2(\u,\,0)$ as a function of $\u$,
while the right  and top  axes respectively show  $C_{\ell}(0)$ as a function of $\ell$. 
In addition to the total foregrounds, the individual components namely 
Point Source Clustering (PSC), Point Source   Poisson (PSP)and  Galactic Synchrotron Emission (GSE).
The Galactic Fre-Free (GFF) and the  Extra-Galactic  Free-Free  (EGFF) components  are relatively much  weaker, have not been  shown but they have in total foreground predictions. }
\label{fig:foregrounds}
\end{centering}
\end{figure}

Figure \ref{fig:foregrounds} shows the total expected  sky signal  assuming  
that the brightest source in the field has a flux of $S_{c}= 1\, \rm Jy$. 
The predictions are shown for Phase II, and  ${F}_2(\u,0)$   can be scaled 
by a factor of $6$ to obtain the predictions for Phase I. The baseline range 
$U\le 10$ is dominated by the synchrotron radiation,  whereas $10 \le U \le 300$ 
is dominated by the clustering of the point sources and $ U \ge 300$ is dominated by the 
Poisson contribution. 
The contributions from the Galactic and extra-galactic free-free
emission are considerably  smaller across the entire $U$ range. 
We find that the total foreground contribution to each visibiity is around 
 $10^4-10^5$ times larger than the \HI signal.

\begin{figure*}
 \vskip1cm \centerline{{ \includegraphics[width=7.cm,height=5.cm
     ]{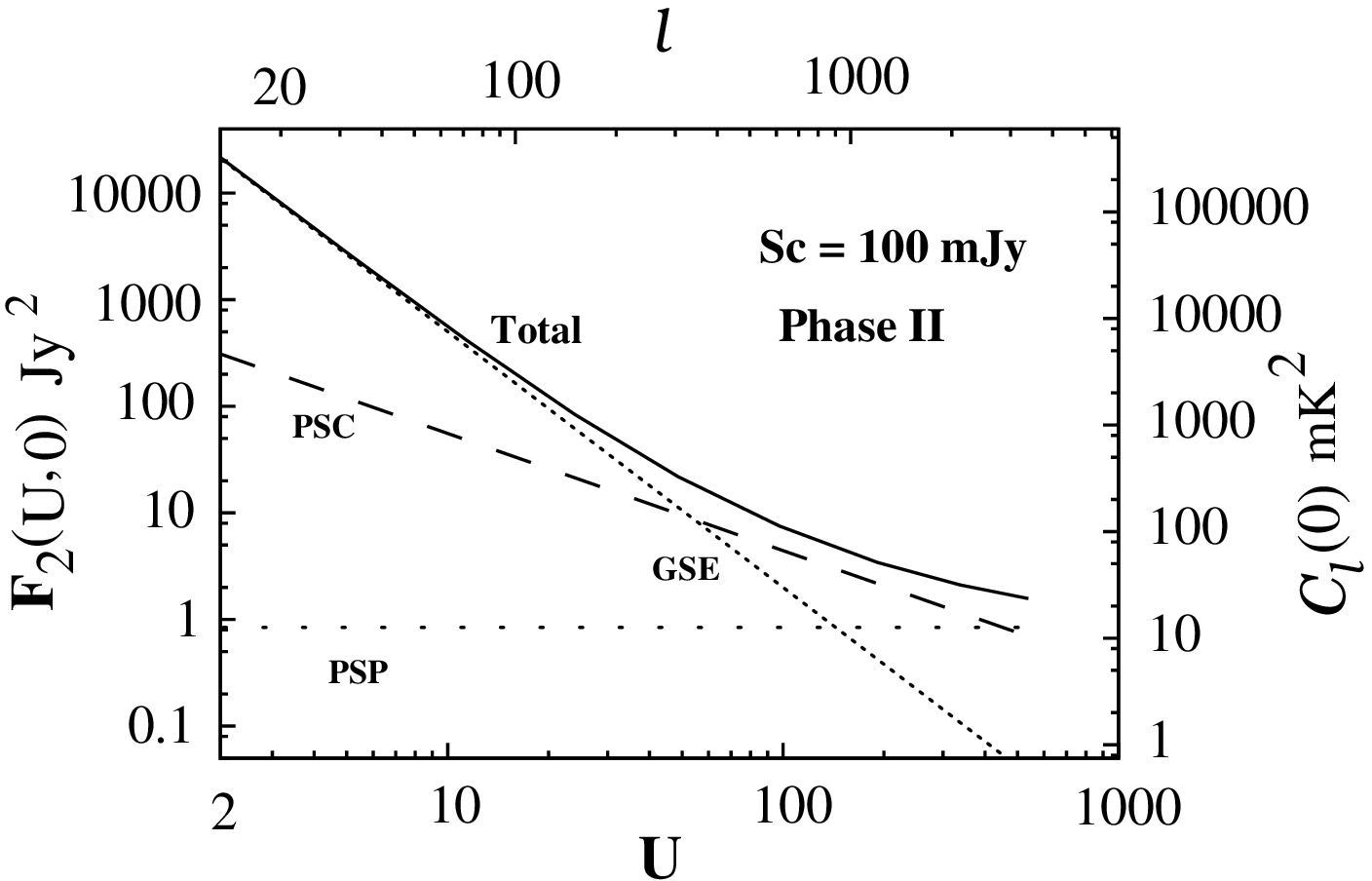}}
    \hskip.2cm { \includegraphics[width=7.cm,height=5.cm
      ]{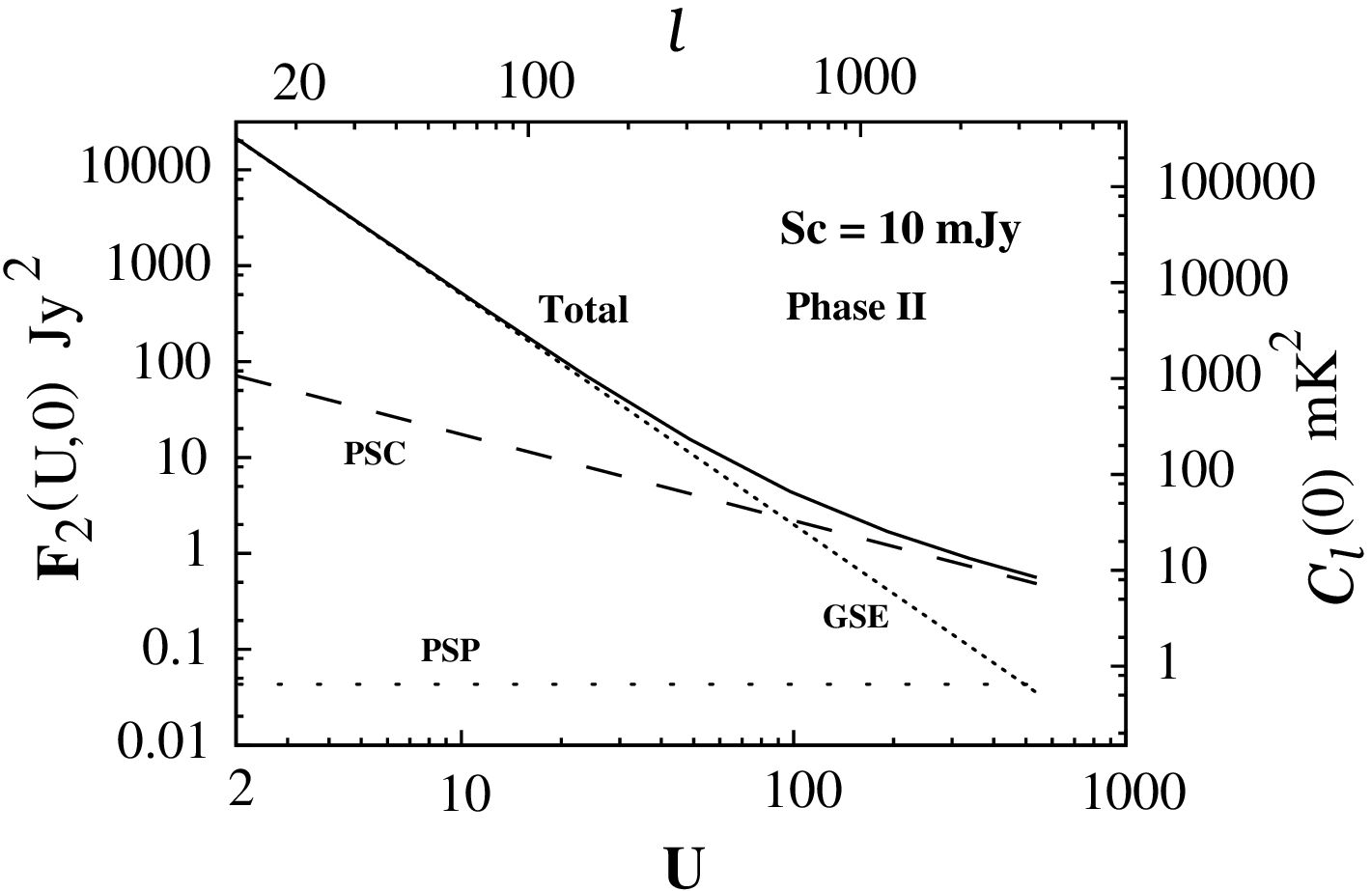}} }
 \caption{Same as Figure \ref{fig:foregrounds} except that 
we assume that the point sources brighter than 
$S_{c}=100 \,\rm mJy$( left panel), and $S_{c}=10\, \rm
   {mJy}$ (right panel) have been identified and subtracted out from the data. 
The GFF and EGFF components, which are relatively much  weaker, have not been shown.}

\label{fig:foregrounds1} 
\end{figure*}

It is very important to correctly identify the point sources and subtract these out 
at a high level of precision ($\sim 10 -100 \, {\rm mJy}$)  in order to detect the \HI signal 
(Ali et al. 2008, Bowman et al., 2009; Bernardi et al., 2011;  Pindor et al., 2011; Ghosh et al., 2012). 
 Here we   assume that sources with flux density   $S \ge S_c$   are visually 
identified and  perfectly  subtracted out  from the data.  The left and right panels of 
Figure \ref{fig:foregrounds1}   shows the  foreground predictions for  $S_{c}= 100 \,{\rm mJy}$ 
and $ 10 \, {\rm  mJy}$ respectively.  The Galactic Synchrotron
radiation  is the most dominant component at $\u \le 50$ (i.e. $\theta \ge 34^{'}$) and
 $\le 100$ (i.e. $\theta \ge 17^{'}$) for $S_{c}= 100 \,\rm {mJy}$ and $ 10 \,\rm {mJy}$ 
respectively. The point source clustering component dominates at larger baselines 
or small angular scales.  The  Poisson contribution falls faster than the  
clustering contribution as $S_c$ is reduced, and it is sub-dominant at all $U$.

The  confusion limit  is predicted to be $\sim 175 \,\rm {mJy}$ for ORT (Phase II).
 However, we do not propose  to identify and subtract  point sources  using 
one dimensional ORT images. We  plan to use  existing $325 \,\rm {MHz}$ source  catalogues
(e.g. The Westerbork Northern Sky Survey (WENSS); Rengelink et al. 1997) or GMRT observations
to identify point sources in the  ORT field of view and subtract their contribution for
the ORT visibility data.  The WENSS survey has a  thresold flux density of  $18 \,\rm {mJy}$, 
whereas the deepest GMRT observation (Sirithia et al. 2009)   has achieved a thresold 
flux density of $0.27\,\rm {mJy}$  at this frequency. 

The residual foregrounds, after point source subtractions, are still $\sim 10^{4}$ times
the \HI signal. As mentioned earlier, we expect ${F}_2(U,\Delta \nu)$ to have a smooth 
$\Delta \nu$ dependence and remain correlated across $\Delta \nu \sim 5\, {\rm MHZ}$ whereas the \HI signal 
is expected to decorrelate within this frequency interval. It is thus, in principle, possible to use
the distinctly different $\Delta \nu$ dependence to separate the \HI signal
from the  foregrounds. This foreground removal technique has been demonstrated to work 
in $610 \, {\rm MHz}$ GMRT observations where it was possible to completely remove the 
foregrounds  so that the residuals were consistent with the cosmological \HI signal and noise 
(Ghosh et al. 2011b). We propose to use a similar technique for foreground removal from the ORT 
data.

\section{Summary and conclusions}
The ORT is currently being upgraded to operate  as a radio-interferometer.
The upgrade is being carried out in two different stages
 with two nearly independent systems, namely {\bf{Phase I}} and
 {\bf{Phase II}}, being expected at the end of the upgrade.  We
have  briefly discussed  these two phases and the relevant parameters
 are presented in Table ~\ref{tab:array}. The telescope has a nominal frequency 
of $326.5 \, {\rm MHz}$ which corresponds to the \HI signal from the  redshift 
$z =3.35 $. 
 
Phases I and II respectively cover the angular multipole range $80 \le \ell \le 3100$ 
and $10 \le \ell \le 3500$, which correspond to the  Fourier modes in the range 
$1.2 \times 10^{-2} \le \kpr \le 5.0 \times 10^{-1} \,{\rm {Mpc}^{-1}}$ 
and $2.0 \times 10^{-3} \le \kpr \le 5.4 \times 10^{-1} \,{\rm {Mpc}^{-1}}$ 
for the 3D \HI power spectrum (Figure ~\ref{fig:pk.eps}). We see that Phase I and II 
are both sensitive to the BAO feature which has the first peak at 
$k=4.5\times 10^{-2} \,{\rm {Mpc}^{-1}}$. The successive oscillations also are  
well within the $k$ range that will be probed.

We have made detailed predictions for both, the \HI signal  and the foregrounds
 expected in Phase I and Phase II. The foregrounds, we find, are dominated by 
the Galactic synchrotron emission at large angular scales whereas the contribution 
from the clustering of point sources dominates at small angular scales.  It is very 
important to correctly identify the point sources and subtract 
them from the data. We find that the Galactic synchrotron emission dominates 
at $\ell \le 630$ and the  contribution from the clustering of point sources dominates 
at $\ell > 630$ if we assume that it is possible to identify and subtract all the 
point sources brighter than $S_c=10 \, {\rm mJy}$ (Figure  \ref{fig:foregrounds1}).  
The foreground contribution to the individual visibilities is predicted to be 
around $10^4-10^5$ times larger than the \HI signal (Figure \ref{fig:signalnoise}). 
Foreground removal is a big  challenge for detecting the \HI signal. 

The \HI signal at a fixed angular multipole $\ell$ but at  two different 
frequencies $\nu$ and $\nu + \Delta \nu$, we find, decorrelates rapidly as
$\Delta \nu$ is increased  (Figure \ref{fig:kappa}).  The \HI signal 
is found to be totally decorrelated for $\Delta \nu \ge 3 \, {\rm MHz}$ for 
the entire $\ell$ range of our interest (Figure   \ref{fig:kappa.5}). 
In contrast the foregrounds originate from continuum sources, and we expect the 
foregrounds to remain correlated across  $\Delta \nu \sim  3 \, {\rm MHz}$.  We propose 
to use this property to extract the \HI signal from the foregrounds. 

We have investigated the SNR for detecting the \HI signal under the  assumption  
that it is possible to completely remove the foregrounds. For both Phases I and II, the SNR peaks around the baseline $U \sim 100$ which corresponds to the angular 
multipole $\ell \sim 630$ (Figure \ref{fig:signaltonoiseratio_u}).  We see that 
for Phase I, a $3 \sigma$ detection is possible  with $\sim 4,000 \, {\rm hrs}$ of 
observation and a $5 \sigma$ detection is possible with $\sim 5,700 \, {\rm hrs}$ of 
observation. A $3 \sigma$ detection is possible with $\sim 1,000 \, {\rm hrs}$ of
 observation with Phase II, and  a   detection  better than $> 5 \sigma$ is 
possible  with  $\sim 2,000 \, {\rm hrs}$ of observation. 

The present paper primarily introduces the ORT as an instrument for exploring the 
high redshift cosmological \HI signal, and presents the expected signal and foreground 
contributions. Preliminary  SNR estimates have been presented,  and these  have  
been used to estimate  the  observing time required to  detect the \HI signal. 
In subsequent work  we plan to perform a more rigorous analysis  of power 
spectrum and parameter estimation, and also address the possibility of 
detecting  the  BAO feature. \\\\\\\\

\section*{Acknowledgment}
The authors would like to thank Jayaram N. Chengalur and  C.R. Subrahmanya
for  involving  and  motivating us in this project. The authors would also like to thank  
P.K. Manoharan and Visweshwar Ram Marthi for their help  with many  technical details of the 
ORT system. The authors also thank all the above, and  Abhik Ghosh, 
Jasjeet S. Bagla, Peeyush Prasad  and Shiv K. Sethi for useful discussions. SSA
would like to acknowledge C.T.S, I.I.T. Kharagpur for 
the use of its facilities and the support by DST, India, under Project No. SR/FTP/PS-
088/2010. SSA would also like to thank the authorities of
the IUCAA, Pune, India for providing the Visiting Associateship programme.

\end{document}